\documentclass[a4paper,12pt]{article}
\usepackage{amsmath,amssymb}
\usepackage{slashed}
\usepackage[pdftex]{hyperref}
\usepackage{cite}
\usepackage{color}
\usepackage{graphicx}

\makeatletter

\newcommand{\nn}{\nonumber \\}

\def\del{\partial}

\def\GN{G_{\rm HS}}
\makeatletter
\g@addto@macro\bfseries{\boldmath}
\makeatother

\def\half{{\frac{1}{2}}}
\def\thalf{{\tfrac{1}{2}}}

\def\unit{{1\kern-.65ex {\rm l}}}
\def\1{{1\kern-.65ex {\rm l}}}

\def\Im{\mathop{\mathrm{Im}}\nolimits}

\def\sign{\mathop{\mathrm{sign}}\nolimits}

\def\tr{\mathop{\mathrm{tr}}\nolimits}

\let\ev=\bracket
\let\Ev=\Bracket

\def\ft{{\widetilde{f}}}

\def\cF{{\cal F}}

\def\cI{{\cal I}}

\def\cN{{\cal N}}
\def\cO{{\cal O}}

\def\bbZ{{\mathbb{Z}}}

\newcommand{\eq} {equation}
\newcommand{\eqa} {eqnarray}
\newcommand{\NN} {\mbox {$\nonumber$}}

\newcount\hour \newcount\minute
\hour=\time \divide \hour by 60
\minute=\time
\def\now{%
\ifnum \hour<13
  \ifnum \hour=0 \advance \hour by 12 \number\hour:\else \number\hour:\fi%
     \ifnum \minute<10 0\fi%
     \number\minute%
\ A.M.%
\else \advance \hour by -12 \number\hour:%
  \ifnum \minute<10 0\fi%
  \number\minute%
  \ P.M.%
\fi%
}

\makeatother

\begin{document}

% format
\baselineskip=18pt  % a la harvmac
\numberwithin{equation}{section}  % make eq labels (sec.num)
%\allowdisplaybreaks  % allow page breaks in displayed eqs

%% print date, time and filename 
%\pagestyle{myheadings}
%\markright{{\tt \jobname.tex} -- \today{} \now}

%%%%%%%%%%%%%%%%%%%%%%%%%%%%%%%%%%%%%%%%%%%
%%%        TITLE BEGINS HERE
%%%%%%%%%%%%%%%%%%%%%%%%%%%%%%%%%%%%%%%%%%%

% ========== title (paper version, a la harvmac) begins here ==========

\thispagestyle{empty}

% Report number
\vspace*{-2cm} 
\begin{flushright}
%{\tt arXiv:yymm.nnnn}\\
HRI/ST/1502, YITP-15-24
\end{flushright}

% title, authors, affiliation
\vspace*{2.5cm} 
\begin{center}
 {\LARGE ABJ Theory in the Higher Spin Limit}\\
 \vspace*{1.7cm}
 Shinji Hirano$^{1,2}$,
 Masazumi Honda$^{3,4}$,\let\thefootnote\relax\footnote{$^4$ The current affiliation: Department of Particle Physics and Astrophysics, Weizmann Institute of Science, Rehovot 7610001, Israel}
 Kazumi Okuyama$^5$,
 and Masaki Shigemori$^{6,7}$\\
 \vspace*{1.0cm} 
 $^1$ School of Physics and Mandelstam Institute for Theoretical Physics,\\
 $^2$ DST-NRF Centre of Excellence in Mathematical and Statistical Sciences (CoE-MaSS), \\
  University of the Witwatersrand,\\
  WITS 2050, Johannesburg, South Africa\\[1ex]
 $^3$ Harish-Chandra Research Institute,\\
Chhatnag Road, Jhusi, Allahabad 211019, India\\[1ex]
 $^5$ Department of Physics,
Shinshu University, Matsumoto 390-8621, Japan\\[1ex]
 $^6$ Yukawa Institute for Theoretical Physics, Kyoto University,\\
Kitashirakawa Oiwakecho, Sakyo-ku, Kyoto 606-8502 Japan\\[1ex]
$^7$
Hakubi Center, Kyoto University,\\
Yoshida-Ushinomiya-cho, Sakyo-ku, Kyoto 606-8501, Japan

\end{center}
\vspace*{1.5cm}

% abstract
\noindent
We study the conjecture made by Chang, Minwalla, Sharma, and Yin on the
duality between the ${\cal N}=6$ Vasiliev higher spin theory on AdS$_4$
and the ${\cal N}=6$ Chern-Simons-matter theory, so-called ABJ theory, 
with gauge group $U(N)\times U(N+M)$. Building on our earlier results on
the ABJ partition function, we develop the systematic $1/M$ expansion,
corresponding to the weak coupling expansion in the higher spin theory,
and compare the leading $1/M$ correction, with our proposed
prescription, to the one-loop free energy of the ${\cal N}=6$ Vasiliev
theory. We find an agreement between the two sides up to an ambiguity
that appears in the bulk one-loop calculation.

\newpage
\setcounter{page}{1} % don't number title page
\setcounter{footnote}{0}

% ========== title (paper version, a la harvmac) ends here ==========

%%%%%%%%%%%%%%%%%%%%%%%%%%%%%%%%%%%%%%%%%%%
%%%           TITLE ENDS HERE
%%%%%%%%%%%%%%%%%%%%%%%%%%%%%%%%%%%%%%%%%%%

%\tableofcontents
%\printindex

%%%%%%%%%%%%%%%%%%%%%%%%%%%%%%%%%%%%%%%%%%%
%%%        MAIN TEXT BEGINS HERE
%%%%%%%%%%%%%%%%%%%%%%%%%%%%%%%%%%%%%%%%%%%

\section{Introduction}

It has long been speculated that string theory in the high energy limit
$E\sqrt{\alpha'}\to\infty$ undergoes drastic reduction of degrees of
freedom due presumably to enhanced symmetries associated with an
infinite number of massless fields which appear in this limit
\cite{Gross:1988ue, Atick:1988si}. This is the extremity of stringy
regime and may reveal what string theory truly is. The infinite number
of massless fields are higher spin fields, and the high energy limit of
string theory may thus yield higher spin (HS) theory. String theory
might then be realized as the symmetry broken phase of HS theory where
the mass scale $1/\sqrt{\alpha'}$ is dynamically generated.

Higher spin theory has generated a great deal of interest recently. This
goes back to the old work of Vasiliev \cite{Fradkin:1987ks,
Vasiliev:1999ba} who constructed interacting theories of massless higher
spin fields that successfully included gravity, i.e., a spin-2
field. The crucial idea was to consider HS theories on de Sitter (dS) or
anti-de Sitter (AdS) space, instead of Minkowski space, in order to
evade no-go theorems concerning massless higher spin fields
\cite{No-GoTheorems}.  Years later, Klebanov and Polyakov
\cite{Klebanov:2002ja} made the important conjecture that the HS theory
on AdS$_4$ space is dual to the $O(N)$ vector model (VM) at critical
points. Substantial and highly nontrivial evidence for the HS/VM duality
was later provided by Giombi and Yin who demonstrated that 3-point
functions of conserved higher spin currents agree on both sides
\cite{Giombi:2009wh}. This conjecture and its generalizations were
further tested successfully at one loop of the HS theory for the vector
models at both UV and IR fixed points \cite{Giombi:2013fka,
Giombi:2013yva, Tseytlin:2013jya, Giombi:2014iua}.  Meanwhile, the
collective field method was applied to the vector models, elucidating
how the HS theory can be directly reconstructed from the VM as well as
providing a new perspective on the origin of the duality as a gauge
phenomenon \cite{Jevicki:2014mfa, Koch:2010cy}.  It should also be noted
that, pioneered by Gaberdiel and Gopakumar, tremendous progress has been
made in the study of the duality between HS theories on AdS$_3$ and
minimal CFT$_2$'s due to the relative simplicity in lower dimensionality
\cite{Gaberdiel:2010pz, Gaberdiel:2010ar, Creutzig:2011fe,
Gutperle:2011kf, Chang:2011mz}.

String theory on AdS space in the limit $\sqrt{\alpha'}/R_{\rm
AdS}\to\infty$ may provide a concrete example in which one can probe the
symmetric phase of string theory in the high energy limit and study its
connection to HS theory.\footnote{In the case of the HS theory on
AdS$_3$ with ${\cal N}=4$ supersymmetries it was shown via the AdS/CFT
correspondence that the HS theory describes a closed subsector in the
symmetric phase of the type IIB string theory on $AdS_3\times S^3\times
T^4$ in the high energy limit \cite{Gaberdiel:2014cha}.} Via the AdS/CFT
correspondence, the limit may also give us the vector model dual to the
HS theory.  Indeed, such an example was suggested by Chang, Minwalla,
Sharma, and Yin (CMSY) \cite{Chang:2012kt} who proposed the HS limit of
AdS$_4$/CFT$_3$ with ${\cal N}=6$ supersymmetries (SUSY), the version
conjectured by Aharony, Bergman, and Jafferis (ABJ)
\cite{Aharony:2008gk} that generalized their earlier work with Maldacena
(ABJM) \cite{Aharony:2008ug}. The gravity theory is M-theory on
$AdS_4\times S^7/\bbZ_k$ with the 3-form field turned on, $C_3\propto
M$, and the dual field theory is the ${\cal N}=6$ $U(N)_k\times
U(N+M)_{-k}$ Chern-Simons-matter (CSM) theory, called the ABJ theory, where $k$ and $-k$ are
the Chern-Simons levels for the two gauge groups. At large $k$, the
M-theory circle of radius $R_{11}= 1/k$ shrinks and  M-theory reduces
to type IIA string theory on $AdS_4\times \mathbb{CP}^3$ with the NSNS
2-form turned on, $B_2 \propto {M\over k}-{1\over 2}$ \cite{Aharony:2008gk, Aharony:2009fc, 
Bergman:2009zh}.  The ingredient crucial
to the HS/VM duality is the presence of the $B_2$ that, in particular,
provides $U(M)$ vectors in the dual field theory. The HS limit proposed
by CMSY is
\begin{equation}
 M,\,\, |k|\,\,\longrightarrow\,\, \infty
 \quad
 {\rm with}\quad t\equiv {M\over |k|}\quad{\rm and}\quad N\quad{\rm finite}
\label{CMSYlimit}
\end{equation}
which is conjectured to be the ${\cal N}=6$ $U(N)$ Vasiliev theory,
constructed by CMSY and Sezgin-Sundell \cite{Sezgin:1998gg}, where the
Newton constant $\GN$ of the HS theory is proportional to
$1/M$\,\,\footnote{In CMSY, the Newton constant $\GN$ was identified with
${1\over M+N}$. However, as we will see below, the finite $M$
corrections instead suggest that the identification $\GN\propto {1\over
M}$ works better.}  and the parity-violating (PV) phase $\theta_0=\pi
t/2$.  This is, in fact, the high energy limit of type IIA string
theory, since the string length is large, $\sqrt{\alpha'}/R_{\rm AdS} \sim
(k/N)^{1/4}\to\infty$. 
As a comparison, let us consider type IIB string theory on $AdS_5\times
S^5$. If we take the $\sqrt{\alpha'}/R_{\rm AdS}^{\rm IIB}\to\infty$
limit, the 't Hooft coupling $\lambda\to 0$ and the dual field theory,
${\cal N}=4$ super Yang-Mills (SYM) theory, becomes free.\footnote{It
should be noted that there has been significant progress in the study of
the free field limit of AdS$_5$/CFT$_4$ \cite{Gopakumar:2003ns}.} This
is in contrast with the ABJ theory which remains nontrivial in the high
energy limit \eqref{CMSYlimit}.

Therefore, the ABJ theory in the HS limit is an ideal setup to
study the high energy regime of string theory and elucidate its
non-trivial dynamics.  In this paper we study the HS limit of CMSY by
(1) developing the systematic $1/M$ expansion of the free energy of the
ABJ theory, (2) calculating the one-loop free energy of the ${\cal N}=6$
HS theory, and (3) subjecting the results to a one-loop test.

The free energy or the partition function of the ABJ(M) theory has been
studied extensively over the last few years thanks to the localization
technique \cite{Kapustin:2009kz} which drastically simplifies path
integrals of supersymmetric gauge theories \cite{Hama:2011ea,
Pestun:2007rz}.  Inspired by the seminal work of Drukker, Marino, and
Putrov \cite{Marino:2011nm} and, in good part, with the use of the
elegant Fermi gas approach developed by Marino and Putrov
\cite{Marino:2011eh}, a great deal about the ABJ(M) partition function has
been uncovered, in particular, at large $N$, both in perturbative
\cite{Fuji:2011km, Marino:2011eh} and nonperturbative expansions
\cite{Hatsuda:2013gj, Grassi:2014uua}. There has also been significant
progress in the study of Wilson loops in the ABJ(M) theory
\cite{Klemm:2012ii} as well as the partition functions of more general
Chern-Simons-matter theories \cite{Marino:2012az}.
However, the ABJ partition function in the HS limit \eqref{CMSYlimit}
has not been much investigated in the literature.  In the current paper,
building on our earlier work \cite{Awata:2012jb, Honda:2013pea}, we
develop a systematic procedure to compute a large $M$ expansion of the
partition function and start exploring the highly stringy regime of the
HS/ABJ duality at finite $N$.  The HS limit can alternatively be
extracted from the conifold expansion developed in
\cite{Drukker:2010nc}, but our approach has the advantage of directly
giving the $1/M$ expansion.\footnote{We thank Marcos Mari\~no for
pointing out to us the use of the conifold expansion for the HS limit.}

To compare the $1/M$ expansion of the ABJ free energy with that of the
HS free energy, an obstacle is the lack of the action for the
Vasiliev theory from which to extract a weak coupling
expansion\footnote{Although there are some propositions about actions of
the Vasiliev theory
\cite{Boulanger:2011dd,Boulanger:2012bj,Doroud:2011xs,Vasiliev:1988sa},
it is not obvious to compute tree level free energy from these actions.
}. In this paper, following Refs.~\cite{Giombi:2013fka,Giombi:2014iua},
we circumvent this problem by computing the one-loop free energy, which
can be computed without the action as long as we know the spectrum, and
by comparing it with the ABJ free energy. For the technical reason, however, the calculation is performed only in the regime $t\ll 1$ and with the help of the result in \cite{Maldacena:2012sf} we infer the form of the one-loop free energy for generic $t$.

 The organization of this paper is as follows: In section
\ref{sec:main_results} we summarize our claim and the main results on
the HS and ABJ free energy and the correspondence between the two
sides. In section \ref{sec:boundary} we review the integral
representation, sometimes referred to as ``mirror description'' of the
ABJ partition function, using which we analyze the free energy in the HS
limit and develop a systematic $1/M$ expansion. Some of the technical
details in section \ref{sec:boundary} are provided in Appendices
\ref{app:expand_S} and \ref{app:mat_int}\@. In section \ref{sec:bulk} we
calculate the one-loop free energy of ${\cal N}=6$ Vasiliev HS theory.
We close our paper with discussions in section \ref{sec:discussions}.

\section{The main results}
\label{sec:main_results}

We first summarize our claim and the main results on the correspondence
between the ${\cal N}=6$ HS and ABJ free energies in the limit
\eqref{CMSYlimit} with $1/M$ corrections.

\begin{figure}[htbp]
\begin{quote}
 \begin{center}
 \includegraphics[height=4cm]{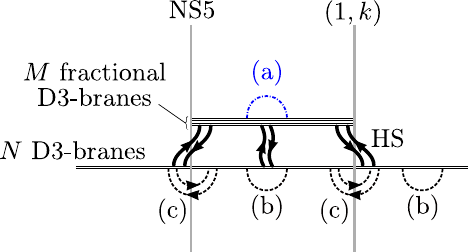}
 \caption{\label{fig:brane_construction} \sl 
The open-string interpretation of the field content of the ABJ theory in the
type IIB UV description.  $N$ D3-branes are intersecting with an
NS5-brane and with a $(1,k)$ 5-brane, and wrap the horizontal direction
which is periodically identified.  $M$ fractional D3-branes 
partially wrap the horizontal direction, ending on the 5-branes.
(For more detail about the brane configuration, see
\cite{Aharony:2008ug,Aharony:2008gk}.)  The open strings stretching
between D3-branes represent fields in the ABJ theory.  To obtain the fields
relevant for the duality to higher spin (HS) theory, we must remove the
open strings related to the $U(M)$ CS theory ((a), blue dashed-dotted line).  
The HS degrees of freedom are dual to combinations of $U(M)$ vectors (thick black lines), 
$U(N)$ adjoints (b) and $U(N)\times U(N)$ bi-fundamentals (c) (black dashed lines).} 
 \end{center}
\end{quote}
\end{figure}

Higher spin theories are dual to vector models.  Our working assumption
is that the vector degrees of freedom dual to the ${\cal N}=6$ HS theory
are massless open strings stretched between $N$ regular and $M$
fractional D3-branes in the type IIB frame of the (UV-completed) ABJ
theory; see figure \ref{fig:brane_construction}.  Since the ABJ theory has a
$U(N)\times U(N+M)$ adjoint and $(\bar{N}, N+M)$ bi-fundamentals with
their conjugates, in addition to the $U(M)$ vectors which are expected to be dual to the higher spin fields, 
we have non-vector degrees of freedom, i.e., (a) the $U(M)$ adjoint, (b) $U(N)\times U(N)$ adjoints,
(c) the $(\bar{N}, N)$ bi-fundamentals and their conjugates.  Note that (b)
and (c) give the same matter content as that appears in the $U(N)_k\times
U(N)_{-k}$ ABJM theory.

Since the $U(M)$ adjoint fields are clearly unwanted degrees of freedom, they have to be removed in the HS/ABJ duality. 
We thus propose that the partition function $Z_{\rm HS}(\GN,\theta_0,N)$ of the ${\cal N}=6$ $U(N)$ Vasiliev HS theory, normalized by the $U(N)$ volume, can be extracted from that of the $U(N)_k\times U(N+M)_{-k}$ ABJ theory, $Z_{\rm ABJ}(N,N+M)_k$, by the quotient\footnote{We revise the proposal in the previous version of our paper,
\begin{align}
Z_{\rm HS}(\GN,\theta_0;N)
 ={1\over {\rm Vol}\left(U(N)\right)}{|Z_{\rm ABJ}(N,N+M)_k|\over Z_{\rm CS}(M)_k\, Z_{\rm ABJM}(N)_k}
\end{align}
which we believe was incorrect.}
\begin{align}
{Z_{\rm HS}(\GN,\theta_0;N)\over {\rm Vol}\left(U(N)\right)}
 ={|Z_{\rm ABJ}(N,N+M)_k|\over Z_{\rm CS}(M)_k}
\label{proposal}
\end{align}
with the identification of the parameters{\footnote{\label{honda}More recently, one of the authors determined the constant $\gamma$ to be $\gamma={2\over \pi}$ by computing the two point function of the stress-energy tensor \cite{Honda:2015sxa}. }
\begin{align}
\GN= {\gamma \over M}{\pi t\over\sin(\pi t)}\qquad\quad\mbox{and}\quad\qquad
\theta_0={\pi t \over 2}\ ,
\label{ParametersId}
\end{align}
where $\gamma$ is a constant that cannot be fixed by the analysis of the
current paper, $t=M/|k|$ as defined in \eqref{CMSYlimit}, and $Z_{\rm
CS}(M)_k$ is the partition function of the ${\cal N}=2$ $U(M)$ Chern-Simons theory at level $k$. 
%The quotient by these factors is to remove contributions from the non-vector degrees of freedom, i.e., $Z_{\rm CS}(M)_k$ and $Z_{\rm ABJM}(N)_k$ corresponding, respectively, to the $U(M)$ adjoint and the $U(N)\times U(N)$ adjoint and bi-fundamentals.

As indicated in figure \ref{fig:brane_construction}, the (massless) open
strings involved in \eqref{proposal} are $U(M)$ vectors and $U(N)\times
U(N)$ bi-fundamentals and adjoints. 
%
%% Hirano san's original
% The way these open strings are
% related to the HS fields is as follows: 
% The $U(M)_{-k}$ vectors,
% consisting of $U(N+M)_{-k}$ adjoints and $U(M)_{-k}\times U(N)_{k}$
% bi-fundamentals, combine themselves with the $U(N)_{-k}\times U(N)_k$
% bi-fundamentals, open strings (c), to form $U(M)_{-k}\times U(N)_k$ and
% $U(M)_{-k}\times U(N)_{-k}$ bi-fundamentals which are further combined
% with their conjugates to form $U(N)_k$ and $U(N)_{-k}$ adjoints. These
% correspond to the HS fields with pure, as opposed to mixed, boundary
% conditions, whereas the $U(N)_k$ and $U(N)_{-k}$ adjoints, open strings
% (b), correspond to spin 1 fields with the mixed boundary condition. (The
% latter would have been absent if the $U(N)$ symmetries were not gauged.)
%
The HS fields, which are $U(N)$ adjoints, arise by connecting these open
strings as follows.  Among the open strings, there are two types of
$U(M)_{-k}$ vectors, namely (i) the $U(M)_{-k}\times U(N)_{-k}$
bi-fundamentals which are contained in the $U(N+M)_{-k}$ adjoint and
represented in Fig.\ \ref{fig:brane_construction} by the middle pair of
black thick arrows, and (ii) the $U(M)_{-k}\times U(N)_{k}$
bi-fundamentals which are represented in Fig.\
\ref{fig:brane_construction} by the pairs of black thick arrows on the
right and left.  Each of these $U(M)_{-k}$-vector strings can be
connected with the $U(N)_{-k}\times U(N)_k$ bi-fundamentals, open
strings (c), to form (i) $U(M)_{-k}\times U(N)_k$ bi-fundamentals and
(ii) $U(M)_{-k}\times U(N)_{-k}$ bi-fundamentals.  The latter
bi-fundamental strings can be further connected with their conjugates on
their $U(M)_{-k}$ endpoints to form (i) $U(N)_k$ and (ii) $U(N)_{-k}$
adjoints.  These $U(N)$ adjoints correspond to the HS fields with pure,
as opposed to mixed, boundary conditions.  On the other hand, the
$U(N)_k$ and $U(N)_{-k}$ adjoints represented by open strings (b)
correspond to spin 1 fields with the mixed boundary condition. (The
latter would have been absent if the $U(N)$ symmetries were not gauged.)

The identification of the Newton constant $\GN$ in \eqref{ParametersId}
can be inferred from the $1/M$ expansion \eqref{F_expl} of the ABJ free energy in which 
$1/M$ systematically appears in the combination $\GN$.
%$F_{\rm HS}(\GN,\theta_0,N)\equiv -\log Z_{\rm HS}(\GN,\theta_0;N)$. Namely, with the identification \eqref{ParametersId}, the $1/M$ expansion of the ABJ free energy, which is the content of section \ref{sec:boundary}, implies the following $\GN$ expansion of the HS free energy:
The proposal \eqref{proposal} then predicts the HS free energy, $F_{\rm HS}\equiv-\log Z_{\rm HS}$, to be\footnote{With the large $M$ expansion we develop in section
\ref{sec:boundary}, one can in principle compute the expansion to
arbitrary finite order. In Eq.~\eqref{F_expl}, we present the explicit
expansion up to order $\GN^4\propto 1/M^4$ terms.}  
\if{\begin{align}
F_{\rm HS}(\GN,\theta_0,N)\equiv&-\log\left[{Z_{\rm HS}(\GN,\theta_0,N)\over{\rm Vol}\left(U(N)\right)}\right]
\nonumber\\
=&{N^2\sin(\pi t)\over g^2(\pi t)}\left[{2{\rm Im}\left[{\rm Li}_2\left(i\tan\theta_0\right)\right]\over \sin(2\theta_0)}-{2\theta_0\ln\tan\theta_0\over\sin(2\theta_0)}\right]
-{N^2\over 2}\ln\left[{N\sin(\pi t)\over 2\pi g^2(\pi t)}\right]\label{HSFreeEnergy}\\
&-\frac{  \left(2 N^2-1\right) (3 \cos (4\theta_0 )+1) }{48} {g^2(\pi t)\over\sin (\pi  t )}+{\cal O}(\GN^4)\ ,
\nonumber
\end{align}
}\fi
\begin{align}
\begin{split}
 F_{\rm HS}(\GN,\theta_0,N)&={\gamma N\over \GN}\,{2\,\cI(\theta_0)\over\sin(2\theta_0)}
 +{N^2\over 2}\ln\!\left({2\gamma\over \pi\GN}\right)
 -{N^2\over 2}\log\left(\sin^2(2\theta_0)\right)\\
 &\qquad
 -(2 N^2-1) (3 \cos (4\theta_0 )+1) {N \GN\over 48\gamma}
 +{\cal O}(\GN^2)
\end{split}
\label{HSFreeEnergy}
\end{align}
where
\begin{align}
 \cI(x) \equiv - \int_0^x dy\, \log\tan y
 ={\Im[{\rm Li}_2(i\tan x)]-x \log\tan x}
 =\cI\Bigl({\pi\over 2}-x\Bigr).
 \label{I_def}
\end{align}
It is worth
emphasizing that the Newton constant $\GN$ agrees with the one suggested by
the computation of three point functions of higher spin currents for
non-supersymmetric theories which is an independent and a completely
different analysis\cite{Aharony:2012nh}. 
Furthermore, as remarked in footnote \ref{honda}, the constant $\gamma$ has been recently determined to be $\gamma=2/\pi$ in \cite{Honda:2015sxa} from the two point function of the stress-energy tensor.

The proposal \eqref{proposal} was motivated in part to respect the invariance under the duality
\begin{align}
 M\,\leftrightarrow\, |k|-M\,,\qquad 
 k\,\leftrightarrow\, -k\ .
\label{KGSduality}
\end{align}
which can be expressed in terms of the HS parameter as
\begin{align}
 \theta_0\,\to\, {\pi\over 2}-\theta_0\ .
\label{KGSduality:t}
\end{align}
In the case of the ABJ theory this is known as the
Giveon-Kutasov-Seiberg duality under which the partition function
$Z_{\rm ABJ}(N,N+M)_k$ is invariant \cite{Giveon:2008zn, Aharony:2008gk}.  For the CS partition
function $Z_{\rm CS}(M)_k$, this is nothing but the level-rank duality. Note that the Newton constant $\GN$ in \eqref{ParametersId} is a duality invariant.  

The HS free energy \eqref{HSFreeEnergy} has a few favorable features: (1) The
leading $1/\GN$ term is linear in $M$, as opposed to $M^2$ as would be
expected from the $U(M)$ vector degrees of freedom, and the dependence
on the PV phase $\theta_0$ is qualitatively similar to that of the
${\cal N}=2$ theory in \cite{Aharony:2012ns} which exhibits the
invariance under $\theta_0\leftrightarrow {\pi\over
2}-\theta_0$. (2) The leading $1/M$ correction, the first logarithmic term
in \eqref{HSFreeEnergy}, is consistent with the one-loop free energy of the
${\cal N}=6$ HS theory whose contribution comes solely from the $U(N)$ gauge
fields, as calculated in section \ref{sec:bulk}, up to the ambiguity of the constant
$\gamma$.

Finally, the presence of the third term $-{N^2\over 2}\log(\sin^2(2\theta_0))$ in \eqref{HSFreeEnergy} may call for a further explanation. This is a part of the HS one-loop contribution and diverges logarithmically as the PV phase $\theta_0$ is switched off or takes the maximal value $\pi/2$.\footnote{In fact, the first term  in \eqref{HSFreeEnergy} which is the classical contribution also diverges logarithmically as $\theta_0\to 0$ or $\pi/2$. With the lack of full understanding of the HS theory action, it is not clear how this singularity should be interpreted.} Although this might look like an unpleasant result, it can be argued that this indeed precisely agrees with the $\tilde{\lambda}$-dependent factor in the anomalous dimension eq.(A.5) of \cite{Maldacena:2012sf} predicted from HS symmetry considerations. 
We will make a more detailed discussion on this point later in Section \ref{sec:discussions}.

\section{The boundary side: ABJ theory}
\label{sec:boundary}

In this section, we study the HS limit of the partition function of the ABJ
theory and develop a systematic way to derive its large $M$ expansion.
The expansion can be explicitly worked out any finite order in
principle.  In the next section, we will use the 1-loop part of the
expansion for comparison with the bulk Vasiliev theory.

\subsection{The ABJ partition function}

The partition function of the $U(N_1)_k\times U(N_2)_{-k}$ ABJ theory on
$S^3$ has been written in the matrix model form \cite{Kapustin:2009kz,
Hama:2011ea} using the localization technique \cite{Pestun:2007rz}.  The
explicit expression of the partition function is
\begin{align}
Z_{\rm ABJ}(N_1,N_2)_k
 =
 \cN\int\prod_{j=1}^{N_1}{d\mu_j\over 2\pi}\prod_{a=1}^{N_2}{d\nu_a\over 2\pi}
 {\Delta_{\rm sh}(\mu)^2\Delta_{\rm sh}(\nu)^2\over\Delta_{\rm ch}(\mu,\nu)^2}
 e^{{ik\over 4\pi}\left(\sum_{j=1}^{N_1}\mu_j^2-\sum_{a=1}^{N_2}\nu_a^2\right)}\ ,
\label{ABJMM}
\end{align}
where $\Delta_{\rm sh}$ and $\Delta_{\rm ch}$ are the one-loop
determinant of the vector multiplets and the matter multiplets in the
bi-fundamental representation, respectively:
\begin{gather}
\Delta_{\rm sh}(\mu)
=\prod_{1\le j<m\le N_1}
 \!\!\left( 2\sinh{\mu_j-\mu_m\over 2}\right) ,
 \quad
 \Delta_{\rm sh}(\nu)=\prod_{1\le a<b\le N_2}
 \!\!\left( 2\sinh{\nu_a-\nu_b\over 2}\right) ,
\\
\Delta_{\rm ch}(\mu,\nu)
 =\prod_{j=1}^{N_1}\prod_{a=1}^{N_2}\left(2\cosh{\mu_j-\nu_a\over 2}\right)\ .
\end{gather}
Furthermore,
$k\in\mathbb{Z}_{\neq 0}$ is the Chern-Simons level,
while $\cN$ is the normalization factor \cite{Marino:2011nm}
\begin{align}
 \cN \equiv {i^{-{\kappa\over 2}(N_1^2-N_2^2)}\over N_1!\, N_2!}\ ,\qquad
 \kappa \equiv\sign k\ .
\end{align}
Because of the relation
\begin{align}
 Z_{\rm ABJ}(N_2,N_1)_{k}
 =Z_{\rm ABJ}(N_1,N_2)_{-k}
 =Z_{\rm ABJ}(N_1,N_2)_{k}^*
 \ ,
 \label{N1N2swap}
\end{align}
we can assume $N_1\le N_2$ and $k>0$ without loss of generality, as we
will do henceforth.  We set
\begin{align}
 N_1\equiv N,\qquad N_2\equiv N+M,\qquad M\ge 0.
\end{align}
We write $Z_{\rm ABJ}(N_1,N_2)$ also as $Z_{\rm ABJ}(N;M)$.

There are various ways to analyze the ABJ partition function
\eqref{ABJMM}, including the Fermi gas approach \cite{Marino:2011eh,
Matsumoto:2013nya, Honda:2014npa} extensively used in the literature.
However, for the purpose of studying its HS limit, the most convenient
starting point is the ``mirror description'' of the ABJ partition
function found in \cite{Awata:2012jb}, generalizing the mirror
description of the ABJM partition function \cite{Benini:2009qs,
Kapustin:2010xq}.
The ``mirror description'' of the ABJ partition function is as
follows:
\begin{align}
 Z_{\rm ABJ}(N;M)_k=
 i^{- N(N+M-1)}2^{-N}k^{-N}q^{{1\over 6}M(M^2-1)}Z_{\rm CS}(M)_k \Psi(N;M)_k,
 \label{AHS}
\end{align}
where
\begin{align}
 Z_{\rm CS}(M)_k
&=
 q^{-{1\over 12}M(M^2-1)}
 k^{-{M\over 2}}
 \prod_{j=1}^{M-1}\left(2\sin{\pi j\over k}\right)^{M-j}
 \label{ZCS}
\end{align}
is the partition function for the $U(M)_k$ CS theory
and we defined the quantity\footnote{Note that $\Psi$ defined in \eqref{Psi_def} is
different from the one in \cite{Awata:2012jb} by the inclusion of the
factor $(-1)^{\half N(N-1)}$.}
\begin{align}
\Psi(N;M)_k&\equiv
 {(-1)^{\half N(N-1)}\over N_1!}\prod_{j=1}^{N}\left[{-1\over 2\pi i}
 \int_{C} {\pi \, ds_j\over \sin(\pi s_j)}\right]
 \prod_{j=1}^{N}{(q^{s_j+1})_{M}\over (-q^{s_j+1})_{M}}
 \prod_{1\le j<m\le N}
 {(1-q^{s_m-s_j})^2\over (1+q^{s_m-s_j})^2}.
\label{Psi_def}
\end{align}
In the above, we defined
\begin{align}
 q\equiv e^{-{2\pi i\over k}},\label{q_def}
\end{align}
and $(a)_n=(a;q)_n \equiv \prod_{j=0}^{n-1} (1-aq^j)$ is the $q$-Pochhammer symbol.  The contour of
integration in \eqref{Psi_def} is $C=\left[-i\infty+\eta, +i
\infty+\eta\right]$ with the constant $\eta$ chosen to lie in the
following range:
\begin{align}
 \begin{cases}
  -M-1< \eta < 0 &  \qquad (k\ge 2M)\\
  -{k\over 2}-1< \eta < -{k\over 2}-M &  \qquad (M\le k\le 2M)\\
 \end{cases}
\label{etaPrescription}
\end{align}

In \cite{Awata:2012jb}, various consistency checks of the expression
\eqref{AHS} were performed: (i) agreement of the perturbative expansion
with the original matrix integral \eqref{ABJMM}, (ii) vanishing of the
partition function for $k<M$, in accord with the prediction
\cite{Aharony:2008gk} that there must be no SCFT in this range, and
(iii) invariance under the Giveon-Kutasov-Seiberg duality
\eqref{KGSduality}. Later, the expression \eqref{AHS} was derived in
\cite{Honda:2013pea} directly from the matrix integral \eqref{ABJMM}
using the Cauchy-Vandermonde formula.

\subsection{The large $M$ expansion}

We would like to develop a formulation to evaluate the ABJ partition
function in the HS limit \eqref{CMSYlimit}.  The expression \eqref{AHS}
is especially suitable for that purpose, since the number of integrals
$N$ is fixed in the HS limit.  To begin with, let us rewrite
\eqref{Psi_def} in the following way \cite{Honda:2014npa}:
\begin{align}
 \Psi(N;M)_k
  &= {1\over N!}
 \left[\prod_{j=1}^{N} \int_{-\infty}^\infty  dx_j\right]
  e^{\sum_{j=1}^{N}f(x_j)}
  \prod_{j<m}^{N}\tanh^2{\pi(x_j-x_m)\over k},
 \label{Psi_ito_f}
\end{align}
where we did the following change of variables
\begin{align}
 s_j=-{M+1\over 2}+ix_j,\qquad j=1,\dots,N,
\end{align}
and also defined
\begin{align}
 f(x,k,t)=\sum_{m=-{M-1\over 2}}^{M-1\over 2}
 \log\tanh {\pi (x+im)\over k}-R(x),\label{def_f(x)}
\end{align}
with
\begin{align}
 R(x)=
 \begin{cases}
  \log(2\cosh(\pi x)) & \qquad (M=2p:\text{even}),\\
  \log(2\sinh(\pi x)) & \qquad (M=2p-1:\text{odd}).\\
 \end{cases}\label{def_R(x)}
\end{align}
In \eqref{def_f(x)}, the summation over $m$ is done in steps of one;
namely, $m=-{M-1\over 2},-{M-1\over 2}+1,\dots,{M-1\over 2}-1,{M-1\over
2}$, whether $M$ is even or odd.  It is easy to show that the
integration contour for $x_j$ in \eqref{Psi_ito_f} corresponds to
choosing $\eta$ correctly in the range \eqref{etaPrescription}, and that
$x=0$ is the critical point of the function $f(x)$ for both even and odd
$M$.  Therefore, the strategy is to expand $f(x)$ around $x=0$ and carry
out the integration by expansion around that point, taking into account
the HS limit \eqref{CMSYlimit}.  It is easy to show that $f(x,k,t)$ is
an even function in $x$.

As we have shown in Appendix \ref{app:expand_S}, using the
Euler-Maclaurin formula, $f(x,k,t)$ can be \emph{formally} rewritten as
\begin{align}
 f(x,k,t)
 &=
 {\cos{2x \partial_t\over k}\over \sinh{\partial_t\over k}}\log\tan{\pi t\over 2},
 \label{expn_f(x)}
\end{align}
in the sense that the formal power expansion of \eqref{expn_f(x)} around
$x=0$ reproduces the formal power expansion of \eqref{def_f(x)}. Namely,
the right hand side gives the asymptotic expansion of $f(x,k,t)$.  Let
us write the expansion of \eqref{expn_f(x)} in $x$ as
\begin{align}
 f(x,k,t)
 \equiv
 \sum_{n=0}^\infty {(-1)^n f_{2n}(k,t)\over (2n)!} {x^{2n}\over k^{2n-1}}
 .
 \label{fn,fnm_def}
\end{align}
Here, the quantities $f_{2n}(k,t)$ are defined as the expansion coefficients and their
explicit expression is given by \eqref{expn_f(x)} as
\begin{align}
 f_{2n}(k,t)&=
 k^{2n-1}
 {({2\partial_t\over k})^{2n}\over \sinh{\partial_t\over k}}\log\tan{\pi t\over 2}
 =\sum_{m=0}^\infty {2^{2n}(2-2^{2m})B_{2m}\over (2m)!\, k^{2m}}
 \partial_t^{2n+2m-1}\log\tan{\pi t\over 2}
 ,\label{f_n}
\end{align}
where $B_n$ are the Bernoulli numbers.  Note that $f_{2n}(k,t)$ is
defined so that its $1/k$ expansion (which is equivalent to the
$1/M$ expansion) starts with an $\cO(k^0)$ term. The $m=0$ term in $f_{0}$
is understood as
\begin{align}
 {1\over \partial_t}\log\tan{\pi t\over 2}
 =\int_0^t dy\log\tan{\pi y\over 2}
 =-{2\over \pi}\,\cI\Bigl({\pi t\over 2}\Bigr),
\end{align}
where $\cI(x)$ was defined in \eqref{I_def}.

If we write down the first few terms of the expansion \eqref{fn,fnm_def}, we have
\begin{align}
 f(x,k,t)&=kf_0(k,t)-{f_2(k,t)\over 2!}{x^2\over k}
           +{f_4(k,t)\over 4!}{x^4\over k^3}-\cdots.
 \label{f_expn_expl}
\end{align}
The first term gives a constant contribution irrelevant for
the $x$ integration, while the $x^2$ term suggests that we define a
new variable $\xi$ by
\begin{align}
 x=k^{1/2}\,\xi,
\end{align}
so that the expansion \eqref{f_expn_expl} now reads
\begin{align}
 f(x,k,t)
 &=
 \sum_{n=0}^\infty {(-1)^n f_{2n}(k,t)\over (2n)!} {\xi^{2n}\over k^{n-1}}
 =kf_0(k,t)-{f_2(k,t)\over 2!}\xi^2+{f_4(k,t)\over 4!}{\xi^4\over k}+\cdots.
\end{align}
Now, the $\xi^2$ term is $\cO(k^0)$ and the higher power terms in $\xi$
are down by powers of $1/k$.  This gives a starting point for the large
$k$ (large $M$) expansion of the integral \eqref{Psi_ito_f}.

In terms of $\xi$, the integral \eqref{Psi_ito_f} can be rewritten as
\begin{align}
 \Psi(N;M)_k
  &= {\pi^{N(N-1)}e^{k N f_0(k,t)}\over N!\,\,k^{{N^2\over 2}-N}}
 \left[\prod_{j=1}^{N} \int_{-\infty}^\infty  d\xi_j\right]
 \Delta(\xi)^2\nn
 &\qquad\times
  \exp\left[\sum_{n=1}^\infty {(-1)^n f_{2n}(k,t)\over (2n)!\, k^{n-1}}\sum_{j=1}^{N}\xi_j^{2n}
 +2\sum_{1\le j<m\le N}\log{\tanh{\pi(\xi_j-\xi_m)\over k^{1/2}}\over {\pi(\xi_j-\xi_m)\over k^{1/2}}}
 \right],
 \label{Psi_ito_xi}
\end{align}
where $\Delta(\xi)$ is the Vandermonde determinant,
\begin{align}
 \Delta(\xi)\equiv
 \prod_{1\le j<m\le N}(\xi_j-\xi_m).
\end{align}
The integral \eqref{Psi_ito_xi} is a standard Hermitian matrix integral and can be
straightforwardly evaluated, regarding the $\xi^2$ term as giving the
propagator and all higher power terms as interactions.  Here we do not
present the detail of the computation but simply write down the
resulting large $M$ expansion:
\begin{align}
 &\cF(N;M)_k
 \equiv -\log \Psi(N;M)_k\notag\\[1ex]
 &=
 {2NM\over \pi t}\,\cI\Bigl({\pi t\over 2}\Bigr)
 +{N^2\over 2}\ln {4M\over \pi t\sin(\pi t)}
 -{N\over 2}\ln{2M^2\over \pi t^2} 
 -\ln G_2(N+1)\notag\\
&\quad
-
 \frac{ N (2 N^2-1)}{48}
\Bigl({\pi t\over M\sin(\pi  t)}\Bigr)
 \left[3 \cos (2 \pi  t )+1\right]
 \nn
&\quad
-
\frac{N^2}{2304}
\Bigl({\pi t\over M\sin(\pi  t)}\Bigr)^{\!2}
 \left[
 (17 N^2+1) \cos (4 \pi  t )
 +4(11 N^2-29) \cos (2 \pi  t ) 
 -157   N^2+211
 \right]\nn
&\quad
-
 \frac{N}{552960}
\Bigl({\pi t\over M\sin(\pi  t)}\Bigr)^{\!3}
\Bigl[
 (674 N^4+250 N^2+201) \cos (6 \pi  t ) \notag\\ &\qquad\quad
 -6 (442 N^4+690 N^2-427) \cos (4 \pi  t )
   +3 (2282N^4+3490 N^2-3635)\cos (2 \pi  t ) \notag\\ &\qquad\quad
 +4348N^4-21940 N^2+12750
 \Bigr]
\notag\\
&\quad
-
\frac{N^2 }{22118400}
\Bigl({\pi t\over M\sin(\pi  t)}\Bigr)^{\!4}
   \Bigl[
 (6223  N^4+8330 N^2+2997) \cos (8 \pi  t )\notag\\ &\qquad\quad
 -8 (3983 N^4+6730 N^2-363) \cos (6 \pi  t )
 +20 (3797 N^4+1870 N^2+1623) \cos (4 \pi  t )\notag\\ &\qquad\quad
 -8 (22249 N^4-44410 N^2+37011) \cos (2 \pi  t)
 -56627 N^4+113630 N^2-18753
 \Bigr]
\notag\\
& \quad
 +\cO(M^{-5}).
 \label{F_expl}
\end{align}
Note that the full ABJ free energy $F_{\rm ABJ}=-\log Z_{\rm ABJ}$ contains
more terms coming from \eqref{AHS}. The computational detail of
\eqref{F_expl} can be found in Appendix \ref{app:mat_int}\@.  Because we
used an asymptotic expansion in evaluating the integral, the large $M$
expansion \eqref{F_expl} is also an asymptotic expansion to be completed
by non-perturbative corrections.

As the last and important remark in this section, we emphasize that as is evident in \eqref{F_expl}, the $1/M$ expansion organizes itself into the $\GN$ expansion, which lead us to the proposal in \eqref{ParametersId}.

%%%%%%%%%%%%%%%%%%%%%%%%%%%%%%%%%%%%%%%%%%%%%%%%%%
%%%%%%%%%%%%%%%%%%%%%%%%%%%%%%%%%%%%%%%%%%%%%%%%%%
%%%%%%%%%%%%%%%%%%%%%%%%%%%%%%%%%%%%%%%%%%%%%%%%%%
\section{The bulk side: $\mathcal{N}=6$ Vasiliev theory}
%%%%%%%%%%%%%%%%%%%%%%%%%%%%%%%%%%%%%%%%%%%%%%%%%%
%%%%%%%%%%%%%%%%%%%%%%%%%%%%%%%%%%%%%%%%%%%%%%%%%%
%%%%%%%%%%%%%%%%%%%%%%%%%%%%%%%%%%%%%%%%%%%%%%%%%%
\label{sec:bulk}

In this section we compute the one-loop free energy of the bulk HS theory dual to the ABJ theory in the higher spin limit \eqref{CMSYlimit}.\footnote{We thank Rajesh Gopakumar for stimulating discussions which motivated us to carry out the calculation in this section.}
It was conjectured in \cite{Chang:2012kt} that
the ABJ theory in the higher spin limit corresponds to
the $\mathcal{N}=6$ parity-violating $U(N)$ Vasiliev theory on $AdS_4$.
The Vasiliev theory has three parameters:
\begin{enumerate}
\item The Newton constant $\GN$ which is proportional to
      $M^{-1}$ at large $M$, as mentioned in the Introduction and section \ref{sec:main_results}.

\item The rank $N$ of the $U(N)$ Chan-Paton factors which is identified with the $N$ of the $U(N)\times U(N+M)$ gauge group of the ABJ
      theory.  

\item The PV phase $\theta_0$ which violates parity and higher spin
      symmetry. As stated in the Introduction, 
      $\theta_0$ is identified with the 't Hooft coupling
      $t$ by $\theta_0=\pi t/2$ 
      \cite{Chang:2012kt,Maldacena:2012sf}.
\end{enumerate}
The partition function of the Vasiliev theory takes the following form
in perturbation theory:
\begin{\eq}
Z_{\rm HS}\equiv e^{-F_{\rm HS}}\qquad\mbox{where}\qquad
 F_{\rm HS}
=\frac{1}{\GN}F_{\rm HS}^{(-1)} +F_{\rm HS}^{(0)} +\GN F_{\rm HS}^{(1)} +\cdots.
\end{\eq}
The free energy $F_{\rm HS}^{(\ell)}$ at
($\ell + 1$)-loops is a function of the PV phase $\theta_0$ and may
receive logarithmic corrections of the form $\GN^{\ell}\log{\GN}$.  The
tree-level free energy $\GN^{-1}F_{\rm HS}^{(-1)}$ is the saddle point action of the Vasiliev theory. 
Although there are some propositions on the actions of
the Vasiliev theory
\cite{Boulanger:2011dd,Boulanger:2012bj,Doroud:2011xs,Vasiliev:1988sa},
it is not obvious to compute the tree level free energy from these actions.  Thus we focus on
the leading correction $F_{\rm HS}^{(0)}$, the
one-loop free energy of the Vasiliev theory.  
The spectrum does not depend on the PV phase $\theta_0$,
and we can compute $F_{\rm HS}^{(0)}$ in the standard manner
\cite{Giombi:2013fka, Giombi:2013yva,Tseytlin:2013jya,Giombi:2014iua, Gaberdiel:2010ar, Gupta:2012he}.

\subsection{The one-loop contribution}

The $\mathcal{N}=6$ Vasiliev theory is constructed from  the so-called $n=6$ extended supersymmetric Vasiliev theory by imposing a set of $SO(6)$ invariant boundary conditions
\cite{Chang:2012kt,Sezgin:2012ag}.
The parity-even $n=6$ Vasiliev theory can have 64 supercharges, but the
boundary conditions and the parity violation reduce the number of supersymmetries to $\mathcal{N}=6$ with 24
supercharges. The spectrum of the $\mathcal{N}=6$ Vasiliev theory
is given by\cite{Chang:2012kt,Sezgin:2012ag}
\begin{itemize}
\item 32 fields for each integer, $s=0,1,\cdots$, and half-integer spin, $s={1\over 2},{3\over 2}, {5\over 2},\dots$ and their associated ghosts with spin $s-1$.

\item All integer and half-integer spin fields with $s\ge 2$ obey the
      so-called $\Delta_+ =s+1$ boundary condition at the $AdS_4$
      boundary, and their associated ghosts have $\Delta_+=s+2$.

\item Half of the spin-0 fields have the $\Delta_+ =1$ boundary condition,
      whereas the other half $\Delta_- =2$.
      
\item Except for one out of thirty-two, the $U(N)$ spin-1 fields have the $\Delta_+ =2$ boundary
      condition and $\Delta_+=3$ for the associated ghosts. The remaining one has the mixed boundary condition, $i\epsilon_{ijk}(\del_jA_k+A_jA_k)+\tan(\pi t)\del_zA_i=0$, with the boundary Chern-Simons term at level $k$, corresponding to the gauging of the $U(N)$ symmetry \cite{Chang:2012kt, Giombi:2013yva}.
      
\item The spin-$0$ ghost field for the spin-$1$ field with the mixed boundary condition has the $\Delta_-=0$ boundary condition \cite{Giombi:2013yva}.

\iffalse
\item 31  of the spin-1 fields have the $\Delta_+ =2$ boundary
      condition, whereas the remaining one satisfies the $\Delta_- =1$
      condition. The $\Delta_-$ spin-1 field is most important in the following and corresponds to a gauge field.
\fi      
      
\end{itemize}
For the technical reason we only deal with the regime $t\ll 1$ and the spin-1 field with the mixed boundary condition in effect has $\Delta=2+{\cal O}(t)\simeq 2$.  We then infer the form of the one-loop free energy for generic $t$ from this data in conjunction with the result of \cite{Maldacena:2012sf}.

We summarize the spectrum in Table \ref{spintable}.
There is a very important point to be stressed: The boundary conditions, as stated here, are only true in the strict large $M$ limit. 
In fact, $\Delta_{\pm}$ are the dimensions of CFT operators dual to higher spin fields and may thus receive $1/M$ corrections which moreover depend on the PV phase $\theta_0$} \cite{Maldacena:2011jn,Maldacena:2012sf}. As we will see, the $1/M$ correction to the $\Delta_-$ spin-0 ghost fields are particularly important and contribute to the one-loop free energy, whereas all the rest of $1/M$ corrections, even if present, have no contributions to one-loop. In Table \ref{spintable} we indicated the ${\cal O}(1/M)$ correction to the $\Delta_-$ spin-0 ghost to emphasize this point.

\begin{table}[htbp]
\def\arraystretch{1.25}
$\bullet$ Spin-$s$ fields
\vspace{.3cm}\\
\begin{tabular}{|c|c|c|c|c|c|c|}
\hline
spin & 0 & 0 & 1 & 1 (gauge)& $s\ge 2$ & $s=\mathbb{Z}_{\ge 0}+{1\over 2}$ \\ \hline
no. of fields & 16 & 16 & 31 & 1 & 32 & 32\\ \hline
boundary cond. & $\Delta_+=1$ & $\Delta_-=2$ & $\Delta_+=2$ & $\Delta\simeq 2$ (mixed) & $\Delta_+=s+1$ & $\Delta_+=s+1$ \\
\hline
\end{tabular}
\vspace{.3cm}\\
$\bullet$ Spin-$(s-1)$ ghosts
\vspace{.3cm}\\
\begin{tabular}{|c|c|c|c|c|c|c|}
\hline
spin & \mbox{ }N/A\mbox{ } & \mbox{ }N/A\mbox{ } & 0 & 0 (gauge)& $s-1\ge 1$ & $s-1=\mathbb{Z}_{\ge 0}+{1\over 2}$ \\ \hline
no. of fields & N/A & N/A & 31 & 1 & 32 & 32\\ \hline
boundary cond. & N/A & N/A & $\Delta_+=3$ & $\Delta_-= 0+{c_-(\theta_0)\over M}$ & $\Delta_+=s+2$ & $\Delta_+=s+2$ \\
\hline
\end{tabular}
\caption{\sl The spectrum of the ${\cal N}=6$ Vasiliev theory (in the regime $t\ll 1$) labeled by
spin, number of fields, and boundary conditions and associated ghosts. Note, in particular,
the ${\cal O}(1/M)$ correction to the $\Delta_-$ spin-0 ghost for the spin-1 gauge
field, where $c_-(\theta_0)$ is known up to a numerical constant. The dimension of other
fields also receives $\cO(1/M)$ corrections which, however, do not contribute to the one-loop free energy.
As mentioned above, the spin 1 field with the mixed boundary condition has $\Delta=2+{\cal O}(t)\simeq 2$ in the regime $t\ll 1$.} \label{spintable}
\end{table}

We can now write down the bulk one-loop partition function. Taking into account the $U(N)$ Chan-Paton factors, it reads
\begin{\eq}
e^{-F_{\rm HS}^{(0)}}
=\Biggl[ Z_{0,\Delta_+}^{16}Z_{0,\Delta_-}^{16} Z_{1,\Delta_+}^{31} Z_{1,\Delta}
           \prod_{s=2}^\infty Z_{s,\Delta_+}^{32} \prod_{s=0}^\infty Z_{s+\frac{1}{2},\Delta_+}^{32} \Biggr]^{N^2},
\label{eq:1loop}
\end{\eq} 
where $Z_{s,\Delta_\pm}$ is the partition function for a field with spin $s$ and the boundary condition $\Delta_{\pm}$
and can be expressed in terms of functional
determinants of symmetric transverse traceless (STT) tensors in $AdS_4$ \cite{Gupta:2012he,
Giombi:2013fka,Tseytlin:2013jya,Giombi:2014iua}:\footnote{In the unit $R_{\rm AdS}=1$.  } 
\begin{\eq}
Z_{s,\Delta_\pm}
=\begin{cases} 
\displaystyle
\left[  \frac{\det_{s-1, \Delta_\pm}^{\rm STT}[-\nabla^2 +(s+1)(s-1)]}{\det_{s, \Delta_\pm}^{\rm STT}[-\nabla^2 +(s+1)(s-2)-s]}   
\right]^{1/2}  &\quad {\rm for}\ s\in{\mathbb{Z}_{\geq 0}}\\[3ex]
\displaystyle
 \left[  \frac{\det_{s, \Delta_\pm}^{\rm STT}
[-\slashed{\nabla}^2 +(s-1/2)^2 ]}
                  {\det_{s-1, \Delta_\pm}^{\rm STT}[-\slashed{\nabla}^2 +(s+1/2)^2 ]}   \right]^{1/4} &\quad {\rm for}\ s\in{\mathbb{Z}_{\geq 0}} +\dfrac{1}{2}
\end{cases} \ ,
\label{spinspart}
\end{\eq} 
with the understanding that 
\begin{\eq}
{\rm det}_{s}^{\rm STT}[ \cdots ] =1\qquad {\rm for}\qquad s<0 \ .
\end{\eq}
$Z_{1,\Delta}$ is the partition function for the spin-1 gauge field with the mixed boundary condition in the regime $t\ll 1$, corresponding effectively to $\Delta\simeq \Delta_+=2$, and its associated ghost with the $\Delta_-$ boundary condition, and a similar one-loop determinant formula  holds for $Z_{1,\Delta}$.
The spin-$(s-1)$ determinants in \eqref{spinspart} are the contributions from the gauge
fixing ghosts. These determinants can be explicitly computed by applying the techniques developed
in \cite{Camporesi:1993mz, Camporesi:1992wn,Camporesi:1992tm}.  
To proceed, we first simplify \eqref{eq:1loop} by using the result of Giombi and Klebanov for the type-A Vasiliev theory \cite{Giombi:2013fka},
\begin{\eq}
Z_{\rm type\,A}= \prod_{s=0}^\infty Z_{s,\Delta_+} =1 .
\end{\eq}
Dividing \eqref{eq:1loop} by $\left(Z_{\rm type \,A}\right)^{32N^2}$ yields
\begin{\eq}
e^{-F_{\rm HS}^{(0)}}
=\Biggl[ \left( \frac{Z_{0,\Delta_-}}{Z_{0,\Delta_+}} \right)^{\!\!16}\,  \frac{Z_{1,\Delta} }{ Z_{1,\Delta_+}}  
 \prod_{s\in\mathbb{Z}_{\geq 0}+{1\over 2}} Z_{s,\Delta_+}^{32} \Biggr]^{N^2} \ .
 \label{Fmassaged}
\end{\eq} 
Thus the bosonic contribution to the one-loop free energy could come only
from the spin-0 and spin-1 fields. This simplifies the calculation.

For the convenience of the subsequent calculations we introduce 
\begin{\eq}
F_{(\Delta ,s)} 
=\begin{cases}
\frac{1}{2} \log{ {\rm det}_{s}^{\rm STT}\bigl[-\nabla^2 +\left( \Delta -\frac{3}{2}\right)^2 -s-\frac{9}{4} \bigr]} & {\rm for}\ s\in\mathbb{Z} \\[1ex]
\frac{1}{2} \log{ {\rm det}_{s}^{\rm STT}\bigl[-\slashed{\nabla}^2 +\left( \Delta -\frac{3}{2}\right)^2 \bigr]}  & {\rm for}\ s\in\mathbb{Z}+\frac{1}{2}
\end{cases} 
\end{\eq}
which has been computed by Camporesi and Higuchi \cite{Camporesi:1993mz, Camporesi:1992wn,Camporesi:1992tm} and is given in terms of the spectral zeta function 
\begin{\eq}
F_{(\Delta ,s)} 
=- \frac{1}{2}\zeta_{(\Delta ,s)}^\prime (0) -\frac{1}{2}\zeta_{(\Delta ,s)} (0)\, \log{ (\Lambda^2 )}\ , 
\label{Fdeltas}
\end{\eq}
where the spectral zeta function $ \zeta_{(\Delta ,s)} (z)$ is defined by
\begin{\eqa}
&& \zeta_{(\Delta ,s)} (z)  = \frac{8(2s+1)}{3\pi} \int_0^\infty \!du\, \frac{\mu_s (u)}{[ u^2 +(\Delta -3/2)^2 ]^z}\ ,\qquad
\zeta_{(\Delta ,s)}^\prime (z) =\frac{\partial}{\partial z} \zeta_{(\Delta ,s)} (z) \ , \NN\\
&& \mu_s (u) =\frac{\pi u}{16} \Biggl[ u^2 +\left( s+\frac{1}{2}\right)^2 \Biggr] \tanh{(\pi (u+is))} \ .
\end{\eqa}
The parameter $\Lambda$ in \eqref{Fdeltas} is a UV cutoff.  The logarithmic divergence arises
in even dimensions and is related to the conformal
anomaly.  As we will show below, the logarithmic divergence actually cancels out
in the $\mathcal{N}=6$ Vasiliev theory (in a certain regularization
scheme).  Hence the net contribution to the one-loop partition function
comes solely from $\zeta_{(\Delta ,s)}^\prime$.  In particular,
the $\mathcal{O}(\log M)$ correction observed in the ABJ theory comes entirely from the $\Delta_-$ spin-0 ghosts for the spin-1 $U(N)$ gauge fields
and the consequence of the ``induced gauge symmetry'' \cite{Giombi:2013yva}.

%%%%%%%%%%%%%%%%%%%%%%%%%%%%%%%%%%%%%%
%%%%%%%%%%%%%%%%%%%%%%%%%%%%%%%%%%%%%%
\subsection{The bosonic contributions}
%%%%%%%%%%%%%%%%%%%%%%%%%%%%%%%%%%%%%%
%%%%%%%%%%%%%%%%%%%%%%%%%%%%%%%%%%%%%%
We first consider the bosonic part $F_{\rm HS,\,B}^{(0)}$ of the one-loop free energy. As commented on below \eqref{Fmassaged}, there are only contributions from the spin-0 and spin-1 fields. 
Moreover, as it will turn out, it is free of logarithmic divergences.  
For integer spins, the spectral zeta function $\zeta_{(\Delta ,s)}(0)$ has been calculated by Camporesi and Higuchi \cite{Camporesi:1993mz, Giombi:2013fka}:
\begin{\eq}
\zeta_{(\Delta ,s)}(0) =\frac{2s+1}{24}\biggl[  \nu^4 -\biggl( s+\frac{1}{2} \biggr)^2 \left( 2\nu^2 +\frac{1}{6}\right) -\frac{7}{240}  \biggr] \quad
{\rm with}\quad \nu =\Delta -\frac{3}{2} \ .
\end{\eq}
Noting that $\Delta_+-3/2=-(\Delta_--3/2)$, this expression 
implies, due to the invariance under $\nu\rightarrow -\nu$, that 
\begin{\eq}
\zeta_{(\Delta_+ ,s)}(0) = \zeta_{(\Delta_- ,s)}(0) \ .
\end{\eq}
Thus the logarithmic divergence in the bosonic part of the free energy cancel out between the contributions from different boundary conditions,
namely, 
\begin{\eq}
\left. \log{\frac{Z_{0,\Delta_-}}{Z_{0,\Delta_+}}} \right|_{\text{log div}} =0\ ,\qquad
\left. \log{\frac{Z_{1,\Delta}}{Z_{1,\Delta_+}}} \right|_{\text{log div}} =0 \ ,
\end{\eq}
where $\dots|_{\text{log div}}$ means the logarithmically divergent part read off from \eqref{Fdeltas}.

Turning to the finite piece, we first calculate the spin-1 free energy.  Again borrowing the result from
\cite{Camporesi:1993mz, Giombi:2013fka} and paying special attention to the ghost boundary conditions, we have\footnote{To be more precise, there is a contribution from the spin 1 fields, ${1\over 2} \bigl(I_B (\Delta -3/2 ,1) -I_B (\Delta_+ -3/2 ,1) \bigr)$, which, however, is at most of order ${\cal O}(t)$ and negligible for our purpose.}
\begin{\eq}
\log{\frac{Z_{1,\Delta}}{Z_{1,\Delta_+}}}
= \frac{1}{2}
 \bigl(I_B (\Delta_+ -3/2 ,0) -I_B (\Delta_- -3/2 ,0) \bigr) \ ,
\end{\eq}
where
\begin{\eq}
 I_B (\nu ,s ) = \frac{2s+1}{3} \int_0^\nu dx \biggl[ \biggl( s+\frac{1}{2}\biggr)^2 x -x^3 \biggr] \psi (x+1/2) 
\end{\eq}
with $\psi(z)$ being the digamma function.
Here, as emphasized in the discussion of the spectrum, we need special care in dealing with the conformal dimensions $\Delta_\pm$.
Generically, the dimensions $\Delta_\pm$  may receive the finite $M$ corrections, and for the spin-0 ghosts it reads
\begin{\eq}
\Delta_+ = 3 +\frac{c_+(\theta_0)}{M} +\mathcal{O}\left( \frac{1}{M^2} \right) \ ,\qquad\quad 
\Delta_- = 0 +\frac{c_-(\theta_0)}{M} +\mathcal{O}\left( \frac{1}{M^2} \right) \ ,
\end{\eq}
where $c_\pm(\theta_0)$ are functions of the PV phase $\theta_0$.
In fact, it has been shown \cite{Maldacena:2011jn,Maldacena:2012sf} that
the $\mathcal{O}(1/M)$ corrections exist in three-dimensional interacting CFTs with pseudo-higher spin symmetries.
When we take into account the $\mathcal{O}(1/M)$ corrections,
an explicit calculation shows that
\begin{\eq}
I_B (\Delta_+ -3/2 ,0) =\mathcal{O}(M^0 ) \ ,\qquad
I_B (\Delta_- -3/2 ,0) = +\log{\left(M/c_-(\theta_0)\right)} +\mathcal{O}(M^0 )\ ,
\end{\eq}
where the $\mathcal{O}(M^0)$ terms are independent of $c_{\pm}(\theta_0)$.  We thus find that
\begin{\eq}
\log{\frac{Z_{1,\Delta}}{Z_{1,\Delta_+}}}
= -\frac{1}{2}\log\left(M/c_-(\theta_0)\right)+{\cal O}(M^0)\ .
\label{Fspin1}
\end{\eq}
Since there is an unknown numerical constant in $c_-(\theta_0)$, we cannot accurately calculate the ${\cal O}(M^0)$ term.
Similarly, it is straightforward to find the spin-0 free energy as
\begin{\eq}
\log{\frac{Z_{0,\Delta_-}}{Z_{0,\Delta_+}}}
= \frac{1}{2}\Bigl( -I_B (-1/2 ,0) +I_B (1/2 ,0) \Bigr) = \mathcal{O}(M^0 )\ .
\label{Fspin0}
\end{\eq}
Combining \eqref{Fspin1} and \eqref{Fspin0} together, we conclude that the bosonic part of the bulk one-loop free energy is
\begin{\eq}
F_{\rm HS,\,B}^{(0)} =  +\frac{N^2}{2}\log\left(M/c_-(\theta_0)\right)  +\mathcal{O}(M^0 )\ .
\end{\eq}
We will later discuss the form of $c_-(\theta_0)$ in Section \ref{sec:discussions}.

%%%%%%%%%%%%%%%%%%%%%%%%%%%%%%%%%%%%%%
%%%%%%%%%%%%%%%%%%%%%%%%%%%%%%%%%%%%%%
\subsection{The fermionic contributions}
%%%%%%%%%%%%%%%%%%%%%%%%%%%%%%%%%%%%%%
%%%%%%%%%%%%%%%%%%%%%%%%%%%%%%%%%%%%%%
We next consider the fermionic part $F_{\rm HS,\,F}^{(0)}$ of the one-loop free energy.
Again, as it will turn out, it is free of logarithmic divergences. 
Moreover, it has no $\log{M}$ corrections.

We first show the absence of the logarithmic divergences:
For $s\in \mathbb{Z}+1/2$,
we can rewrite the spectral zeta function $\zeta_{(\Delta ,s)} (z)$ as a sum of two terms
\begin{\eq}
\zeta_{(\Delta ,s)} (z)  = \frac{8(2s+1)}{3\pi} (g_1 (\nu ,s ;z) +g_2 (\nu ,s ;z) ) \ ,
\end{\eq}
where 
\begin{\eqa}
&& g_1 (\nu ,s ;z) = \frac{\pi}{16}  \int_0^\infty \! du\, \frac{u}{( u^2 +\nu^2 )^z}  \Biggl[ u^2 +\left( s+\frac{1}{2}\right)^2 \Biggr] ,\NN\\
&& g_2 (\nu ,s ;z) = \frac{\pi}{8}  \int_0^\infty \! du \, \frac{u}{( u^2 +\nu^2 )^z (e^{2\pi u}-1)}  \Biggl[ u^2 +\left( s+\frac{1}{2}\right)^2 \Biggr] .
\end{\eqa}
By explicit calculations, these two terms are given by
\begin{\eq}
g_1 (\nu ,s ;0) = \frac{\pi\nu^2}{64} \Biggl[ \nu^2 -\left( s+\frac{1}{2}\right)^2 \Biggr] \ ,\qquad
g_2 (\nu ,s ;0) 
=\frac{\pi  (20 s (s+1)+7)}{3840} \ .
\label{f1f2}
\end{\eq}
Meanwhile, from \eqref{Fmassaged} and \eqref{Fdeltas}, the logarithmically divergent piece of $F_{\rm HS,\,F}^{(0)}$ is
\begin{\eq}
 -8N^2\biggl[ \zeta_{(3/2,1/2)} (0)
 +\sum_{s\in\mathbb{Z}_{\geq 0}+1/2} \Bigl( \zeta_{(s+1,s)} (0) -\zeta_{(s+2,s-1)} (0) \Bigr) \biggr] 
\log{( \Lambda^2 )} \ .
\end{\eq}
This sum, as it stands, is divergent, and must be  regularized.  
We adopt the regularization used in the analysis \cite{Giombi:2014iua}.\footnote{This regularization can be slightly generalized to:
\begin{\eq}
  \zeta_{(3/2,1/2)} (0)
 +\lim_{\alpha\rightarrow 0} \sum_{s\in\mathbb{Z}_{\geq 0}+1/2} \left( s+x\right)^{-\alpha} \zeta_{(s+1,s)} (0) 
 -\lim_{\alpha\rightarrow 0} \sum_{s\in\mathbb{Z}_{\geq 0}+1/2} \left( s+y\right)^{-\alpha} \zeta_{(s+2,s-1)} (0) \ .
\end{\eq}
One can show that this vanishes so long as $x+y=0$.
}
This yields
\begin{align}
\left. F_{\rm HS,\,F}^{(0)}\right|_{\text{log div}} &= -8N^2 \biggl[ 
 \zeta_{(3/2,1/2)} (0)
 +\lim_{\alpha\rightarrow 0} \sum_{s\in\mathbb{Z}_{\geq 0}+1/2} s^{-\alpha}
\Bigl( \zeta_{(s+1,s)} (0) -\zeta_{(s+2,s-1)} (0) \Bigr)   \biggr] \log{( \Lambda^2 )} \NN\\
&=32\Biggl[ \frac{11}{360}
 +\lim_{\alpha\rightarrow 0} \sum_{s\in\mathbb{Z}_{\geq 0}+1/2}   s^{-\alpha} 
\left( -\frac{5 s^4}{12}+\frac{5 s^2}{24}+\frac{13}{2880} \right) \Biggr]    \log{( \Lambda^2 )}
 =0\ , 
\end{align}
where we used \eqref{f1f2} to find the second line. Thus the fermionic part of the one-loop free energy is also free of logarithmic divergences.

We next evaluate the finite part.  For $s\in\bbZ_{\ge 0}+1/2$, an explicit computation yields
\begin{align}
 \zeta_{(\Delta ,s)}^\prime (0)  
&= -\frac{8(2s+1)}{3\pi} \left( (s+1/2)^2 d_1 +d_3 \right) +I_F (\nu ,s ) \NN\\
& \qquad -\frac{(2s+1)}{72} \nu  \left(-3 \nu ^3+4 \nu ^2+\nu -12 s^2-12 s-3\right) \ ,
\end{align}
where
\begin{\eq}
d_n = \frac{\pi}{8} \int_0^\infty du \,\frac{u^n \log{u^2 }  }{ e^{2\pi u}-1}\ ,\qquad
 I_F (\nu ,s ) = \frac{2s+1}{3} \int_0^\nu dx \Biggl[ \left( s+\frac{1}{2}\right)^2 x -x^3 \Biggr] \psi (x)\ .
\end{\eq}
It is then straightforward to show that
each piece in the finite part is of order $\mathcal{O}(M^0)$, 
\begin{\eq}
 \zeta_{(s+1 ,s)}^\prime (0) =\mathcal{O}(M^0 )  \ ,\qquad\qquad
 \zeta_{(s+2 ,s-1)}^\prime (0) =\mathcal{O}(M^0 )  \ . 
\end{\eq}
Hence the ${\cal O}(\log M)$ contribution is absent in the fermionic free energy, and it is at most of order $\mathcal{O}(M^0)$,
\begin{\eq}
F_{\rm HS,\,F}^{(0)} =  \mathcal{O}(M^0 ) \ .
\end{\eq}

%%%%%%%%%%%%%%%%%%%%%%%%%%%%%%%%%%%%%%
%%%%%%%%%%%%%%%%%%%%%%%%%%%%%%%%%%%%%%
\subsection{The full one-loop free energy}
%%%%%%%%%%%%%%%%%%%%%%%%%%%%%%%%%%%%%%
%%%%%%%%%%%%%%%%%%%%%%%%%%%%%%%%%%%%%%

Altogether, we find the full bulk
one-loop free energy to be
\begin{\eq}
 F_{\rm HS}^{(0)}=F_{\rm HS,\,B}^{(0)}+F_{\rm HS,\,F}^{(0)} =+\frac{N^2}{2}\log\left(M/c_-(\theta_0)\right) +\mathcal{O}(M^0 ) \ .
 \label{HSFEsec4}
\end{\eq}
Note that the leading ${\cal O}(\log M)$ contribution comes entirely from the $\Delta_-$ spin-0 ghosts for the spin-1 $U(N)$ gauge fields
and, as in \cite{Giombi:2013yva}, is the consequence of the ``induced gauge symmetry.'' 

The bulk one-loop free energy \eqref{HSFEsec4} is consistent with the $\mathcal{O}(\log M)$ correction to the ABJ free energy  
with the identification \eqref{ParametersId} of the Newton constant 
\begin{align}
\GN={\gamma\over M}{\pi t\over\sin(\pi t)}\ .
\end{align}
We are, however, unable to determine the constant $\gamma$ which requires the precise value of the ${\cal O}(M^0)$ correction.\footnote{Once again, as remarked in footnote \ref{honda}, the constant $\gamma$ has been recently determined to be $\gamma=2/\pi$ by one of the authors in \cite{Honda:2015sxa}.} We will make further comments on $c_-(\theta_0)$ in the one-loop free energy in the next section.

\if{\begin{\eq}
\GN = \frac{c}{N+M+n} ,
\label{eq:Newton}
\end{\eq}
where $c$ is a constant and $n$ is an integer.
In terms of $\GN$, we rewrite the bulk one-loop free energy as
\begin{\eq}
 F_{\rm bulk}^{(0)} =+\frac{N^2}{2}\log{\GN} +\mathcal{O}(\GN^0 ) .
\label{eq:bulk}
\end{\eq}
In the next section, we compare this result with the field theory
result.
}\fi

%%%%%%%%%%%%%%%%%%%%%%%%%%%%%%%%%%%%%%%%%%%%%%%%%%
%%%%%%%%%%%%%%%%%%%%%%%%%%%%%%%%%%%%%%%%%%%%%%%%%%
%%%%%%%%%%%%%%%%%%%%%%%%%%%%%%%%%%%%%%%%%%%%%%%%%%
\section{Discussions}
%%%%%%%%%%%%%%%%%%%%%%%%%%%%%%%%%%%%%%%%%%%%%%%%%%
%%%%%%%%%%%%%%%%%%%%%%%%%%%%%%%%%%%%%%%%%%%%%%%%%%
%%%%%%%%%%%%%%%%%%%%%%%%%%%%%%%%%%%%%%%%%%%%%%%%%%
\label{sec:discussions}

In the last two sections, we have calculated the free energies of the
ABJ theory in the HS limit and the $\mathcal{N}=6$ Vasiliev theory at
one-loop.  We are now ready to discuss the correspondence between the two theories.
However, it is not as straightforward as comparing the free energy of the ABJ theory \eqref{F_expl} and that of the ${\cal N}=6$ HS theory \eqref{HSFEsec4} as they are, 
and it requires some considerations to make the correspondence more precise.

\if{First note that the full ABJ free energy does not directly correspond to the one of the Vasiliev
theory and we should appropriately extract the ``vector model sub sector''
in order to compare the both free energies.  We can easily see this from
the fact that the ABJ free energy scales as $\sim M^2$ while the
Vasiliev free energy is $\mathcal{O}(M)$.  Although the way to extract
vector model sector is well understood for Chern-Simons theories with
fundamental matter \cite{Giombi:2011kc}, it has not been investigated
for theories with bi-fundamental matter, namely our case.  Here we
clarify dictionary of the higher spin AdS/CFT correspondence in terms of
the results on the both sides.  
}\fi
As already mentioned in section \ref{sec:main_results}, the ABJ theory, even in the HS limit \eqref{CMSYlimit}, has more degrees of freedom than necessary to describe the ${\cal N}=6$ HS dual. For instance, the free energy of the ABJ theory in the limit \eqref{CMSYlimit} goes as $M^2$, since the ABJ theory is a theory of $U(M)$ matrices. On the other hand, the free energy of the HS theory is expected to grow as $M$, reflecting the fact that it is dual to a $U(M)$ vector model.
The $M^2$ growth comes from the $U(M)$ part of the $U(N)\times U(N+M)$ CS free energy.
In the case of $U(M)$ CS theory coupled to fundamental matter \cite{Giombi:2011kc}, the ${\cal O}(M)$ growth was extracted by normalizing the CS partition function to be unity, or equivalently, dividing the full partition function by the CS partition function. In our case, however, the situation is more involved, since the gauge group is a product group $U(N)\times U(N+M)$ and the ABJ theory has bi-fundamental matter. 

Here we first recall our proposal made in section \ref{sec:main_results} and then elaborate on it. The proposed correspondence is given in \eqref{proposal}:
\begin{\eq}
\frac{Z_{\rm HS} (\GN, \theta_0; N)}{{\rm Vol}\left(U(N)\right)}
=Z_{\rm vec}(M; N)_k\ ,
\label{eq:vec}
\end{\eq}
where the \lq\lq vector model subsector'' of the  partition function is identified as
\begin{align}
Z_{\rm vec}(M;N)_k=\frac{\left| Z_{\rm ABJ}(N, N+M)_k \right|}
{ Z_{\rm CS}(M)_k }\ .\label{quotients}
\end{align}
In addition to the quotient by the $U(M)$ CS partition function on the RHS, 
the LHS of \eqref{eq:vec} is divided by the $U(N)$ volume, 
${\rm Vol}\left(U(N)\right)=(2\pi )^{\frac{N}{2}(N+1)}/G_2 (N+1)$.
\if{Note that our proposal \eqref{eq:vec} is different in some points from
the case of theories with fundamental matter.  For example, if we
consider the $U(N)_k$ Chern-Simons theory with one complex massless
scalar in fundamental representation, dual to the type-A Vasiliev theory
in the free limit \cite{Klebanov:2002ja}, then the partition function of
the vector model sector is given by a ratio between the full partition
function and $Z_{\rm CS}(N)_k $.  From this fact, one might naively
expect that the vector model sector of the ABJ partition function was
also given by $|Z_{\rm ABJ}(N;M)_k|/Z_{\rm CS}(N+M)_k$.  However, we can
easily see that the ratio $|Z_{\rm ABJ}(N;M)_k|/Z_{\rm CS}(N+M)_k$ does
not correctly reproduce the bulk result \eqref{eq:bulk}.  
}\fi
This is the natural normalization for the bulk $U(N)$ theory.
The main idea behind \eqref{quotients} is to regard the open strings stretched between $N$ regular and $M$ fractional (and $N$ regular) D3-branes as the vector degrees of freedom dual to the HS theory, as illustrated in Figure \ref{fig:brane_construction} for the type IIB brane construction of the ABJ(M) theory. 
Thus the quotient by $Z_{\rm CS}(M)_k$ is to remove contributions from the diagrams that only involve open strings whose both ends are on $M$ fractional D3-branes. 
As quantitative justifications, we note that the free energy $F_{\rm vec}=-\log{Z_{\rm vec}}$ of the vector model
subsector has the following properties:
\begin{enumerate}
\item $F_{\rm vec}$ 
scales as $M\propto\GN^{-1}$ at the leading order in the HS limit (and of order ${\cal O}(N^2)$ when expressed in terms of the bulk `t Hooft coupling $\lambda_{\rm HS}=N\GN$, as it should be for $U(N)$ theory).

\item $F_{\rm vec}$ enjoys the Giveon-Kutasov-Seiberg duality
      \eqref{KGSduality}, namely,
\begin{\eq}
F_{\rm vec} (M;N)_k = F_{\rm vec} (|k|-M;N)_{-k}\ .\label{dualityINV}
\end{\eq}

\item The leading logarithmic correction agrees with the bulk one-loop result \eqref{HSFEsec4},
\begin{\eq}
F_{\rm vec}(M;N)_k=\cdots +\frac{N^2}{2}\log M+\cdots\ .
\end{\eq}

\end{enumerate}
We have already emphasized the importance of the first property. Meanwhile, the second property might look a matter of aesthetics. However, the duality invariance \eqref{dualityINV} ensures the parity symmetry restoration at $\theta_0=0$ and ${\pi\over 2}$ with the identification $\theta_0=\pi t/2$ where $t=M/|k|$, as required by the PV Vasiliev theory \cite{Chang:2012kt}. Had it been the $U(N+M)$ CS partition function $Z_{\rm CS}(N+M)_k$ to be divided in \eqref{quotients}, the duality invariance would not have been respected. 
This vindicates the quotient by the $U(M)$ CS partition function $Z_{\rm CS}(M)_k$ as opposed to $Z_{\rm CS}(N+M)_k$.
Lastly, as already stated in previous sections, the third property implies the agreement between the ABJ and HS theories, provided that the HS Newton constant is identified as 
\begin{align}
\GN={\gamma\over M}{\pi t\over \sin(\pi t)}\quad
 \xrightarrow{~t\to 0~}
 \quad{\gamma\over M}
\end{align}
which agrees with the one suggested in \cite{Aharony:2012nh} for non-supersymmetric theories.
We emphasize once again that the HS Newton constant $\GN$, rather than simply $1/M$, is the expansion parameter that appears in the systematic $1/M$ expansion  \eqref{F_expl} of the ABJ free energy.
%It must be stressed that the quotient by the ABJM partition function $Z_{\rm ABJM}(N)_k$ is  crucial for us to be able to make this identification. To elaborate on this point, we note that \if{First, we easily see that $F_{\rm vec}$ satisfies the condition 4 since the each building block of $F_{\rm vec}$ is invariant under the Seiberg like duality.  To check the other three conditions, we need an explicit expression of $F_{\rm vec}$ in the higher spin limit. By using the weak-coupling expansion of the ABJM free energy (see e.g.\ \cite{Marino:2011nm})}\fi
To this end, we spell out the free energy for the RHS of \eqref{quotients} which lead to the main result \eqref{HSFreeEnergy}: 
\begin{align}
F_{\rm vec}(M;N)_k=&{\rm Re}\left[F_{\rm ABJ}(N, N+M)_k\right] -F_{\rm CS}(M)_k\nonumber\\
=&\, {2NM\over \pi t}\, {\cal I}\Bigl({\pi t\over 2}\Bigr) +{N^2\over 2}\log\left({2\over \pi}{M\sin(\pi t)\over \pi t}\right)
-{N^2\over 2}\log\left(\sin^2(\pi t)\right)\\&
+\log\left({\rm Vol}\left(U(N)\right)\right)
+\mathcal{O}\left(\pi t/(M\sin(\pi t))\right)\ . \nonumber
\end{align}
\if{and
\begin{align}
F_{\rm ABJM}(N)_k 
= N^2 \log{\frac{2M}{\pi t}}+\log{\frac{{\rm Vol}\left(U(N)\right)^2}{(2\pi )^{N^2}}} 
+\mathcal{O}(M^{-2}) \ .\label{F2}
\end{align}
Subtracting \eqref{F2} from \eqref{F1}, we find the free energy for the \lq\lq vector model subsector'' to be
\begin{\eqa}
F_{\rm vec}(M;N)_k
= \frac{2NM}{\pi t}\,\cI\Bigl({\pi t\over 2}\Bigr) 
 -\log\bigl[{\rm Vol}(U(N))\bigr]
-\frac{N^2}{2} 
 \log\biggl({1\over 2\pi}{\frac{M\sin(\pi t)}{\pi t}}\biggr) 
   +\mathcal{O}(M^{-1})\ .
\end{\eqa}
Rewriting this in terms of the Newton constant leads to
\begin{align}
F_{\rm vec}
&= \frac{2cN}{\pi t}\,\cI\Bigl({\pi t\over 2}\Bigr) \GN^{-1}  +\frac{N^2}{2} \log{\GN}
-\frac{N^2}{2} \log{\frac{\sin(\pi t)}{2\pi^2 t}} 
\notag\\
&\qquad\qquad
 -\frac{2N(N+n)}{\pi t}\,\cI\Bigl({\pi t\over 2}\Bigr) -\frac{N^2}{2} \log{c} 
  +\mathcal{O}(\GN ). 
\end{align}
Thus, if we take\footnote{%
$I(t)$ around $t=0$ behaves as $I(t) =(-1+\log{(\pi t/2)})t +\mathcal{O}(t^3 )$.
}
\begin{\eq}
n= -N\ {\rm or}\ 0 ,
\end{\eq}
then $F_{\rm vec}$ satisfies the above four conditions.
}\fi
\if{Several comments are in order.
First, we regard the above four conditions as necessary but not sufficient.
Namely, the conditions do not uniquely determine $F_{\rm vec}$.
Second, division by $Z_{\rm CS}(M)_k$ seems natural
since $Z_{\rm CS}(M)_k$ is factorized in the ABJ partition function $Z_{\rm ABJ}$ \eqref{AHS}
and duality invariant thanks to the level-rank duality.
Finally, division by $Z_{\rm ABJM}$ provides\footnote{%
Note that the full ABJ free energy includes $+(M^2 /2) \log{M}$ and
$+(N^2 /2) \log{M}$ coming from $Z_{\rm CS}(M)_k$ and $\Psi (N;M)_k$, respectively.
} 
the true logarithmic behavior $-(N^2 /2) \log{M}$ for $F_{\rm vec}$, 
which agrees with the bulk 1-loop free energy \eqref{eq:bulk}.
Generically it might be more appropriate  
if we replaced $Z_{\rm ABJM}$ by some quantity
which has $-N^2 \log{M}$ and is invariant under the Seiberg-like duality. 

Possible interpretations.  (Dis)similarity  to
Giombi-Klebanov.  Shift in $N$, Jevicki.  CS sector.
}\fi
As promised, we would like to add more comments on the logarithmic terms in the second line. The first logarithmic term is identified with $+N^2/2\log\left(\GN^{-1}\right)$ up to a numerical constant as in \eqref{HSFreeEnergy}. As noted in the end of Section \ref{sec:main_results}, the second logarithmic term diverges as $t\to 0$ or $1$, and this might look like an unpleasant result. However, we now argue that this is indeed precisely the result predicted in \cite{Maldacena:2012sf} from HS symmetry considerations.\footnote{Two comments are in order: 
(1) Due to the $U(N)$ symmetry and supersymmetries, the spectrum of the ${\cal N}=6$ theory is larger than that assumed in \cite{Maldacena:2012sf}. Thus, strictly speaking, we are pushing the applicability of their results potentially beyond the limits. 
(2) This argument of \cite{Maldacena:2012sf} applies to dimensions of CFT operators dual to higher spin fields. We are, however, applying their result to dimensions of bulk ghosts, even though there are no CFT operators dual to them. It is, however, reasonable to assume that the $O(1/M)$ corrections to the dimensions of spin $(s-1)$-ghosts appear in the same form as those of their associated spin $s$-fields. } To see it, note that comparing these two terms with the HS one-loop result \eqref{HSFEsec4}, we wish to show that
\begin{align}
{c_-(\theta_0)\over M}=\GN\sin^2(\pi t)
\end{align} 
up to a numerical constant. In \cite{Maldacena:2012sf} it was suggested in eq.(A.5) that
\begin{align}
{c_-(\theta_0)\over M}=a\,\GN{\tilde{\lambda}^2\over 1+\tilde{\lambda}^2}
+b\, \GN{\tilde{\lambda}^2\over (1+\tilde{\lambda}^2)^2}
\end{align}
where $a$ and $b$ are unknown constants. Meanwhile, $\tilde{\lambda}$ for the ${\cal N}=6$ theory was conjectured in  \cite{Chang:2012kt} to be 
\begin{align}
\tilde{\lambda}=\tan(2\theta_0)=\tan(\pi t)\ .
\end{align}
Provided that $b=0$ for the ${\cal N}=6$ theory, it indeed yields
\begin{align}
{c_-(\theta_0)\over M}=a\,\GN\sin^2(\pi t)
\end{align}
as we wished.
It should also be noted that from the field theory viewpoint, the ABJ theory is related to the ${\cal N}=3$ $U(N+M)_{-k}$ Chern-Simons-matter theory with $2N$ fundamental hypermultiplets by gauging the $U(N)$ subgroup of the flavour symmetry.  The logarithmic singularity (as well as $+{N^2\over 2}\log M$ term) is nothing but the one which appears in the difference of the free energies of the ABJ and the ${\cal N}=3$ theories and similar to the one in eq.(4.18) of \cite{Giombi:2013yva}.

%we add a few words to the comment made in the end of Section \ref{sec:main_results} concerning the term $-{N^2\over 2}\log\left(\sin^2(\pi t)\right)$ which diverges logarithmically as $t\to 0$ or $1$. In the one-loop calculation of the HS free energy with the boundary Chern-Simons term, a similar singularity was suggested in eq.(7.33) of \cite{Giombi:2013yva}.\footnote{{\color{red}With the identification $\tan(\pi t)=\pi N_{\rm theirs}/(8 k)$ proposed in \cite{Chang:2012kt}, the second term in eq.(7.33) of \cite{Giombi:2013yva} can be read off as $-{N^2\over 4}\log\left(\sin^2(\pi t)\right)$.}} Meanwhile, 

We believe that all indicate our proposal \eqref{eq:vec} and \eqref{quotients} is at work. However, it is worth noting that the \lq\lq vector model subsector'' may be a misnomer, since open strings stretched between $M$ fractional and $N$ regular D3-branes, corresponding to the $U(M)$ vector, do couple with open strings which ends only on $M$ fractional D3-branes, corresponding to the $U(M)$ adjoint. Although the quotients \eqref{quotients} do remove all diagrams that \emph{only} involve the latter degrees of freedom, it is \emph{not} the case that these degrees of freedom do not appear at all in Feynman diagrams. %So the quotients \eqref{quotients} do not completely single out $U(M)$ vectors, and there is no $U(M)$ vector subsector in the strict sense.

%%%%%%%%%%%%%%%%%%%%%%%%%%%%%%%%%%%%%%%%%%%%%%%%%%
%%%%%%%%%%%%%%%%%%%%%%%%%%%%%%%%%%%%%%%%%%%%%%%%%%
%%%%%%%%%%%%%%%%%%%%%%%%%%%%%%%%%%%%%%%%%%%%%%%%%%
\section*{Acknowledgments}
%%%%%%%%%%%%%%%%%%%%%%%%%%%%%%%%%%%%%%%%%%%%%%%%%%
%%%%%%%%%%%%%%%%%%%%%%%%%%%%%%%%%%%%%%%%%%%%%%%%%%
%%%%%%%%%%%%%%%%%%%%%%%%%%%%%%%%%%%%%%%%%%%%%%%%%%

We thank Robert de Mello Koch, Yasuaki Hikida, Antal Jevicki, Rajesh
Gopakumar, Shiraz Minwalla, Sanefumi Moriyama, Keita Nii, Eric
Perlmutter, Mukund Rangamani, Joao Rodrigues, Xi Yin, and Costas Zoubos
for useful discussions.
SH would like to thank the Graduate School of Mathematics, Nagoya university, and the Yukawa Institute for Theoretical Physics for their hospitality at various stages of this work. The work of SH was supported in part by the National Research Foundation of South Africa and DST-NRF Centre of Excellence in Mathematical and Statistical Sciences (CoE-MaSS).
Opinions expressed and conclusions arrived at are those of the author and are not necessarily to be attributed to the NRF or the CoE-MaSS.
The work of KO was supported in part by JSPS Grant-in-Aid for 
Young Scientists (B)  23740178.
MS is grateful to the Weizmann Institute for the stimulating environment
at the ``Black Holes and Quantum Information'' workshop.
The work of MS was supported in part by Grant-in-Aid for Young
Scientists (B) 24740159 from the Japan Society for the Promotion of
Science (JSPS)\@.

\appendix

\section{Formal expansion of $f(x,k,t)$}
\label{app:expand_S}

In this Appendix, we derive the formal expansion \eqref{expn_f(x)} of
the quantity $f(x,k,t)$ defined in \eqref{def_f(x)}.

First, let us do the following trivial rewriting of \eqref{def_f(x)} as
\begin{align}
 f(x,k,t)
 =
 \sum_{m=-{M-1\over 2}}^{M-1\over 2} \!\!\log{\tanh {\pi (x+im)\over k}\over{\pi (x+im)\over k}}
 +\sum_{m=-{M-1\over 2}}^{M-1\over 2} \!\!\log{\pi (x+im)\over k}
 -R(x).\label{f_alt}
\end{align}
The quantity $f_{2n}(k,t)$, which was defined in
\eqref{fn,fnm_def} and can be written as
\begin{align}
 f_{2n}(k,t)=(-1)^n k^{2n-1}\partial^{2n}_x f(x,k,t)|_{x=0},
\end{align}
is computed from the expression \eqref{f_alt} as follows.  First, for
even $M$,\footnote{Recall that the summation is always done in steps of one.}
\begin{align}
 f_{2n}
 &=
 \begin{cases}
 k^{-1}\left[
 \sum\limits_{m=-{M-1\over 2}}^{M-1\over 2} \!\!\!\!
 \log{\tan {\pi m\over k}\over{\pi m\over k}}
 +2\sum\limits_{m=\half}^{M-1\over 2}\log{\pi m\over k}
  -\log 2
  \right]
  &(n=0),
 \\[4ex]
 k^{2n-1}\left[
 \sum\limits_{m=-{M-1\over 2}}^{M-1\over 2} \!\!\!\!
 \partial_m^{2n} \log{\tan {\pi m\over k}\over{\pi m\over k}}
 -2(2n-1)!\sum\limits_{m=\half}^{M-1\over 2}{1\over m^{2n}}
  -(-1)^n{(2\pi)^{2n}(2^{2n}-1)B_{2n}\over 2n}
 \right]
  &(n\ge 1).
 \end{cases}
 \label{f2n_expr_evenM}
\end{align}
Here, we used the relation $\partial_x=-i\partial_m$ and the formula
\cite[eq.\ 1.518.2]{GR}
\begin{align}
 R^{\text{$M$:\,even}}(x)=
 \log (2\cosh(\pi x))=
 \log 2+\sum_{n=1}^{\infty }{(2\pi)^{2n}(2^{2n}-1)B_{2n}\over 2n(2n)!}x^{2n}.
\end{align}
For odd $M$, some care is needed in setting $x=0$, because the
singularity at $x=0$ coming from the $m=0$ term in the second sum of
\eqref{f_alt} cancels against the singularity coming from $R(x)$.  Using
the formula \cite[eq.\ 1.518.1]{GR}
\begin{align}
 R^{\text{$M$:\,odd}}(x)=
 \log (2\sinh(\pi x))=
 \log(2\pi x)+\sum_{n=1}^{\infty }{(2\pi)^{2n}B_{2n}\over 2n(2n)!}x^{2n},
\end{align}
we obtain, for odd $M$,
\begin{align}
 f_{2n}
 &=
 \begin{cases}
 k^{-1}\left[
 \sum\limits_{m=-{M-1\over 2}}^{M-1\over 2} \!\!\!\!
 \log{\tan {\pi m\over k}\over{\pi m\over k}}
 +2\sum\limits_{m=1}^{M-1\over 2}\log{\pi m\over k}
  -\log(2k)
 \right]
  &(n=0),
 \\[4ex]
 k^{2n-1}\left[
 \sum\limits_{m=-{M-1\over 2}}^{M-1\over 2} \!\!\!\!
 \partial_m^{2n} \log{\tan {\pi m\over k}\over{\pi m\over k}}
 -2(2n-1)!\sum\limits_{m=1}^{M-1\over 2}{1\over m^{2n}}
  -(-1)^n{ (2\pi)^{2n}B_{2n}\over 2n}
 \right]
  &(n\ge 1).
 \end{cases}
 \label{f2n_expr_oddM}
\end{align}

Because the summand in the first terms of \eqref{f2n_expr_evenM},
\eqref{f2n_expr_oddM} is regular at $m=0$ thanks to the rewriting
\eqref{f_alt}, it can be safely evaluated using the Euler-Maclaurin formula.
The version of the Euler-Maclaurin formula relevant here is the one
that uses the midpoint trapezoidal rule and is given by (see
e.g.\ \cite{Lyness-Ninham})
\begin{align}
&g\Bigl(a+\half\Bigr)+g\Bigl(a+{3\over 2}\Bigr)+\cdots+ g\Bigl(b-\half\Bigr)
 \notag\\
 &\qquad\qquad
 =\int_a^b dt\, g(t)
 +\sum_{n=1}^w {(2^{-2n+1}-1)B_{2n}\over (2n)!} [g^{(2n-1)}(m)-g^{(2n-1)}(0)]+R_{2w-1},\label{Euler-Maclaurin_midpoint}
\end{align}
where the remainder function is
\begin{align}
 R_w&={(-1)^{w+1}\over w!}\int_0^m dt\, g^{(w+1)}(t)\,\zeta\Bigl(-w,t+{1\over 2}\Bigr)
\end{align}
and $\zeta(s,q)$ is the Hurwitz zeta function.
Generally, $R_w$ does not vanish in the $w\to \infty$ limit and, therefore,
sending $w\to\infty$ and dropping $R_w$ in
\eqref{Euler-Maclaurin_midpoint} gives a non-convergent asymptotic
expansion.

For $n\ge 1$, the second terms of \eqref{f2n_expr_evenM} and \eqref{f2n_expr_oddM}
involve the generalized harmonic number,
\begin{align}
 H^{(r)}_q=\sum_{m=1}^q {1\over m^{r}}.
\end{align}
Its asymptotic expansion for large $q$  is \cite{HarmonicNum}
\begin{align}
 H_q^{(r)} \sim \zeta(r) 
 - {2 q+r + 1\over 2(r - 1) (q + 1)^{r} }
 - {1\over (r - 1)!} \sum_{l=1}^\infty
 {(2l+r-2)! \, B_{2l}\over (2l)!\, (q + 1)^{2l + r - 1}},
 \label{harmonic_num_asympt}
\end{align}
where ``$\sim$'' means an asymptotic expansion and $\zeta(s)$ is the
Riemann zeta function.  By expanding this in $r$ around $r=0$ and collecting the
$\cO(r)$ terms, we obtain the asymptotic expansion
\begin{align}
 \sum_{m=1}^q \log m \sim 
 \half \log(2\pi)-1-q
 +(q+\half)\log(q+1)
 + \sum_{k=1}^\infty  {B_{2 k}\over 2 k (2k-1) (q + 1)^{2 k -1}},\label{logsum_asympt}
\end{align}
which we can use for evaluating the $n=0$ case of \eqref{f2n_expr_evenM}
and \eqref{f2n_expr_oddM}.

Applying the above formulas \eqref{Euler-Maclaurin_midpoint},
\eqref{harmonic_num_asympt} and \eqref{logsum_asympt} to
\eqref{f2n_expr_evenM} and \eqref{f2n_expr_oddM} and massaging the
resulting expression, we obtain the following asymptotic expansion:
\begin{align}
 f_{2n}&\sim
 \begin{cases}
  \displaystyle
 \int_{0}^{t} dy\, \log\tan{\pi y \over 2}
 +2\sum_{l=1}^\infty  {(2^{-2l+1}-1)B_{2l}\over (2l)!}
 {(2\partial_t)^{2l-1}\over k^{2l}}
  \log\tan{\pi t \over 2}
 +\ft_0
  &\qquad (n=0),
  \\[3ex]
  \displaystyle
 2\sum_{l=0}^\infty {(2^{-2l+1}-1)B_{2l}\over (2l)!}
 {(2\partial_t)^{2n+2l-1}\over k^{2l}}
 \log\tan{\pi t\over 2}
 +\ft_{2n}
  &\qquad (n\ge 1),
 \end{cases}\label{f2n_final}
\end{align}
where, for even $M$,
\begin{align}
 k\ft_0
 &=
 2\sum_{l=1}^\infty {(2^{2l-1}-1)B_{2l}\over 2l(2l-1)M^{2l-1}}
 +(2M+1)\,\log\!\left(1+{1\over M}\right)-(M+1)\,\log\!\left(1+{2\over M}\right)
 \notag\\
 &\qquad
 +2\sum_{l=1}^\infty {B_{2l}\over 2l(2l-1)}
 \left[{1\over (M+1)^{2l-1}}-{1\over ({M\over 2}+1)^{2l-1}}
 \right],
 \label{f-tilde_0_oddM}
 \\
 {\ft_{2n}\over k^{2n-1}}&=
 2\sum_{l=0}^\infty {2^{2n}(2^{2l-1}-1)(2n+2l-2)!\,B_{2l}\over (2l)!\, M^{2l+2n-1}}
 +(2n-2)!\left[{2^{2n}(2M+2n+1)\over (M+1)^{2n}}-{M+2n+1\over ({M\over 2}+1)^{2n}}\right]
 \notag\\
 &\qquad
 +2\sum_{l=1}^\infty {(2l+2n-2)!\,B_{2l}\over (2l)!}
 \left[{2^{2n}\over (M+1)^{2l+2n-1}}-{1\over ({M\over 2}+1)^{2l+2n-1}}\right]
 \label{f-tilde_2n_oddM}
\end{align}
while, for odd $M$,
\begin{align}
 k\ft_0
 &=
 2\sum_{l=1}^\infty {(2^{2l-1}-1)B_{2l}\over 2l(2l-1)M^{2l-1}}
 +M\,\log\!\left(1+{1\over M}\right)-1
 +2\sum_{l=1}^\infty {2^{2l-1} B_{2l}\over 2l(2l-1)\, (M+1)^{2l-1}},
 \label{f-tilde_0_evenM}
 \\
 {\ft_{2n}\over k^{2n-1}}&=
 2\sum_{l=0}^\infty {2^{2n}(2^{2l-1}-1)(2n+2l-2)!\,B_{2l}\over (2l)!\, M^{2l+2n-1}}
 \notag\\
 &\qquad
 +(2n-2)!{2^{2n}(M+2n)\over (M+1)^{2n}}
 +2\sum_{l=1}^\infty {2^{2l+2n-1}(2l+2n-2)!\,B_{2l}\over (2l)!\,(M+1)^{2l+2n-1}}
 \label{f-tilde_2n_evenM}
\end{align}
with $n\ge 1$.  Some comments in deriving the expression
\eqref{f2n_final} are in order.  First, the first terms in
\eqref{f2n_expr_evenM}, \eqref{f2n_expr_oddM} were evaluated using the
Euler-Maclaurin formula \eqref{Euler-Maclaurin_midpoint} and formally
dropping the remainder function.  In the resulting integrals, we defined
$y\equiv 2m/k$ and rewrote it in terms of $y$-integrals.  For $n\ge 1$, the
integral can be trivially integrated to give the $l=0$ term in
\eqref{f2n_final}.  Furthermore, we split $\log[(\tan{\pi y\over
2})/({\pi y \over 2})]=\log[\tan({\pi y\over 2})]-\log({\pi y\over 2})$
and put the ones originating from $\log({\pi y\over 2})$ into
$\ft_0,\ft_{2n}$.  Next, the second terms in \eqref{f2n_expr_evenM},
\eqref{f2n_expr_oddM} were evaluated using the asymptotic formulas
\eqref{harmonic_num_asympt}, \eqref{logsum_asympt}.  For odd $M$, there
is no problem in directly applying the these formulas but, for even $M=2p$, we
need to use the following trick,
\begin{align}
    \sum_{m=\half}^{p-\half} \log j
 &= -2p\log 2
 +\sum_{m=1}^{2p} \log m -\sum_{m=1}^{p} \log m
,\\
  \sum_{m=\half}^{p-\half} {1\over m^{2n}}
  &=2^{2n} \sum_{m=1}^{2p} {1\over m^{2n}}
 - \sum_{m=1}^{p} {1\over m^{2n}},
\end{align}
before applying the asymptotic formulas.  The asymptotic formula
\eqref{harmonic_num_asympt} involves the $\zeta$ function which may look
like a nuisance, but it precisely cancels the last (constant) terms in
\eqref{f2n_expr_evenM}, \eqref{f2n_expr_oddM}, due to the identity
\begin{align}
 \zeta(2n)={(-1)^{n+1}(2\pi)^{2n}B_{2n}\over 2(2n)!},\qquad
 n\ge 1.
\end{align}
Similar cancellations happen for the $\log$ terms for $n=0$.

Actually, as we will show below, $\ft_0=\ft_{2n}=0$.  Therefore,
\eqref{f2n_final} actually becomes
\begin{align}
 f_{2n}&\sim
 2\sum_{l=0}^\infty {(2^{-2l+1}-1)B_{2l}\over (2l)!}
 {(2\partial_t)^{2n+2l-1}\over k^{2l}}
 \log\tan{\pi t\over 2}\qquad (n\ge 0),
 \label{f2n_final2}
\end{align}
where it is understood that, for $n=l=0$,
\begin{align}
 {1\over \partial_t}
 \log\tan{\pi t \over 2}
=
 \int_0^{t} dy
 \log\tan{\pi y \over 2}.
\end{align}
Formally carrying out the summation in \eqref{f2n_final2}, we obtain
\begin{align}
 f_{2n}\sim
 {(2\partial_t)^{2n}\over k\sinh{\partial_t\over k}}
 \log\tan{\pi t \over 2}.\label{f2n_final3}
\end{align}
If we substitute the expression \eqref{f2n_final3} into
\eqref{fn,fnm_def} and formally perform  the summation over $n$, we
obtain the expression in the main text, \eqref{expn_f(x)}.

The final result \eqref{f2n_final2} may look like the expression which
we would obtain if we directly applied the Euler-Maclaurin formula
\eqref{Euler-Maclaurin_midpoint} to the original expression
\eqref{def_f(x)}.  However, of course, the Euler-Maclaurin formula does
not work in the presence of a singularity that gives a divergent
integral.  It is only after the above careful treatment of the
singularities as we did above and the delicate cancellation of terms due
to the  presence of the seemingly unwanted function $R(x)$ that we
arrived at the very simple expression \eqref{f2n_final2}.

\subsubsection*{$\bullet$ Proof of $\ft_{2n}=0$}

Let us show that $\ft_{2n}=0$ as mentioned above.  For simplicity, let
us consider the case with odd $M$ and $n\ge 1$.  The relevant expression
is \eqref{f-tilde_2n_evenM}.  First, because $B_0=1$, $B_1=-1/2$ and
$B_{2n+1}=0$ for $n\ge 1$, we can combine the two terms in the second
line to get the following expression:
\begin{align}
 {\ft_{2n}\over k^{2n-1}}&=
 2\sum_{l=0}^\infty {2^{2n}(2^{2l-1}-1)(2n+2l-2)!\,B_{2l}\over (2l)!\, M^{2l+2n-1}}
 +\sum_{l=0}^\infty {(-1)^l\, 2^{l+2n}(l+2n-2)!\,B_{l}\over l!\,(M+1)^{l+2n-1}}.\label{f2n_1st+2nd}
\end{align}
When expanded in  $1/M$, the second term is equal to
\begin{align}
 &\sum_{l=0}^\infty {(-1)^l\,2^{l+2n}(l+2n-2)!\, B_{l}\over l!\,M^{l+2n-1}}
 \sum_{p=0}^\infty (-1)^p {l+2n+p-2\choose p}{1\over M^{p}}
 \notag\\
 &\qquad
 =
 \sum_{q=0}^\infty \sum_{l=0}^q {(-1)^q\, 2^{l+2n}!\, (q+2n-2)!\,B_{l}\,\over M^{q+2n-1}l!\, (q-l)!}
 \qquad (l+p\equiv q)
 \notag\\
 &\qquad
 =
 \sum_{q=0}^\infty 
 {(-1)^q \,2^{2n} (q+2n-2)!\, B_{l}\over M^{q+2n-1}q!}
 \sum_{l=0}^q {q\choose l} 2^{l} B_{l}.\label{f2n_2nd}
\end{align}
Now, recalling the relation between the Bernoulli polynomial $B_n(x)$ and the Bernoulli numbers $B_n$,
\begin{align}
 B_n(x)=\sum_{l=0}^{n} {n\choose l} x^{n-l} B_l,
\end{align}
and also the relation
\begin{align}
 B_n(\thalf)=(2^{1-n}-1)B_n,
\end{align}
we find
\begin{align}
 \sum_{l=0}^q {q\choose l}  2^{l} B_{l}
 =
 2^q \sum_{l=0}^q {q\choose l}  \left(\half\right)^{q-l} B_{l}
 =2^q B_q(\thalf)
 =2^q (2^{1-q}-1)B_q.
\end{align}
Therefore,
\begin{align}
\eqref{f2n_2nd}
 &=
 \sum_{q=0}^\infty 
 {(-1)^q\, 2^{q+2n} (2^{-q+1}-1)(q+2n-2)!\, B_{q}\over q!\, M^{q+2n-1}}.
\end{align}
Because the summand vanishes for $q=1$ and because $B_{2n+1}=0$ for
$n\ge 1$, we can set $q=2l$, $l\ge 0$.
Then this cancels the first term in \eqref{f2n_1st+2nd}. So, we have
shown $\ft_{2n}=0$.

In a quite similar manner, using Bernoulli polynomial/number identities, we can show that $\ft_0=0$ for even $M$ and
$\ft_0=\ft_{2n}=0$ $(n\ge 1)$ for odd $M$.

\section{Evaluation of the matrix integral (\ref{Psi_ito_xi})}
\label{app:mat_int}

In this appendix, we would like to systematically evaluate the integral
\eqref{Psi_ito_xi}, which we write down here again for convenience:
\begin{align}
 \Psi(N;M)_k=e^{-\cF(N;M)_k}
  &= {\pi^{N(N-1)}e^{k N f_0}\over N!\,\,k^{{N^2\over 2}-N}}
 \left[\prod_{j=1}^{N} \int_{-\infty}^\infty  d\xi_j\right]
 \Delta(\xi)^2
\nn
 &\qquad\times
  \exp\left[\sum_{n=1}^\infty {(-1)^n f_{2n}\over (2n)!\, k^{n-1}}
 \sum_{j=1}^N \xi_j^{2n}
 +2\sum_{j<m}\log{\tanh{\pi(\xi_j-\xi_m)\over k^{1/2}}\over {\pi(\xi_j-\xi_m)\over k^{1/2}}}
 \right].
 \label{Psi_ito_xi2}
\end{align}
Note that $\cF$ defined here is different from the full ABJ free energy
$F_{\rm ABJ}=-\log Z_{\rm ABJ}$ which contains more terms coming from
\eqref{AHS}.

Because $f_{2n}=f_{2n}(k,t)=\cO(k^0)$, we can treat the $\xi^2$ term in
the exponential of \eqref{Psi_ito_xi2}
as the propagator and all higher power terms as
interactions, and evaluate the integral perturbatively in a $1/k$
expansion.  The last term in the exponential can be written as
\begin{align}
 \sum_{j<m}\log{\tanh{\pi(\xi_j-\xi_m)\over k^{1/2}}\over {\pi(\xi_j-\xi_m)\over k^{1/2}}}
 &=
 \sum_{n=1}^\infty 
 c_{2n} \Bigl({\pi^2 \over k}\Bigr)^{n}
 \sum_{j<m}  (\xi_i-\xi_j)^{2n}
\end{align}
where we used the relation \cite[eq.\ 1.518.3]{GR}
\begin{align}
 \ln {\tan x\over x}
 &=\sum_{n=1}^\infty c_{2n}x^{2n},\qquad
 c_{2n}= {(-1)^{n+1}(2^{2n-1}-1)2^{2n}B_{2n}\over n(2n)!}.
\end{align}

To avoid clutter, let us use the shorthand notation
\begin{align}
 \prod_{j=1}^N \int_{-\infty }^\infty d\xi_j \equiv \int d^N \xi,\qquad
 \sum_{j=1}^N \xi_j^{n} \equiv \xi^{n},\qquad
 \sum_{1\le j<m\le N} (\xi_j-\xi_m)^{2n}\equiv (\Delta \xi)^{2n}.
\end{align}
First, note that the Gaussian integral of the quadratic term is given by
\begin{align}
 \int d^N \xi \,\,
 \Delta(\xi)^2\,
 e^{-{f_2\over 2} \xi^2}
 = f_2^{-{N^2\over 2}}(2\pi)^{N\over 2}G_2(N+2),
\end{align}
where $G_2(N)$ is the Barnes $G$-function.  For a quantity $\cO(\xi)$,
let us define its expectation value by
\begin{align}
 \ev{\cO} \equiv
 { \int d^N \xi \, \Delta(\xi)^2\, e^{-{f_2\over 2}\xi^{2}}\, \cO
 \over
 \int d^N \xi \, \Delta(\xi)^2\, e^{-{f_2\over 2}\xi^{2}} }.
 \label{jxks17Feb14}
\end{align}
Then the integral \eqref{Psi_ito_xi2} can be written as
\begin{align}
 e^{-\cF(N;M)_k}
  &= 
 {2^{N\over 2}\, G_2(N+1)\,\pi^{N^2-{N\over 2}} e^{k N f_0}
 \over 
 k^{{N^2\over 2}-N}f_2^{N^2\over 2}}
 \notag\\
 &\qquad
 \times
 \Ev{
 \exp\left[{\sum_{n=2}^\infty {(-1)^n f_{2n}\over (2n)! \, k^{2n-1}}\xi^{2n}
 +
 \sum_{n=1}^\infty c_{2n} \Bigl({\pi^2 \over k}\Bigr)^{n} (\Delta \xi)^{2n}}\right]
 },
 \label{F_xi_int}
\end{align}
where we used the relation $G_2(z+1)=\Gamma(z)G_2(z)$.

The above is sufficient for computing $\cF(N;M)_k$ in principle, but the
following observation makes the computation simpler. Note that
$\Delta(\xi)^2$ is nothing but the Fadeev-Popov determinant for going
from the matrix model of an $N\times N$ Hermitian matrix $X$ to the diagonal gauge
where $\xi_j$, $j=1,\dots,N$ are the eigenvalues of $X$.  So, the expectation value of
$\cO$ defined in \eqref{jxks17Feb14} can be written as the expectation
value in a Hermitian matrix model as
\begin{align}
 \ev{\cO}=
 {
 \int d^{N^2}\! X\, e^{-{f_2\over 2}\tr X^2}\,
 \cO
 \over 
 \int d^{N^2}\! X\, e^{-{f_2\over 2}\tr X^2}
 }~,
\end{align}
where $X$ is an $N\times N$ Hermitean matrix.  When going
from the eigenvalue basis in terms of $\xi_j$ back to the
Hermitean matrix model, we do the following replacements in $\cO$:
\begin{align}
 \xi^{2n}&=\sum_{i} \xi_i^{2n}\to \tr X^{2n},\\
 (\Delta \xi)^{2n}
 &=\sum_{i<j}(\xi_i-\xi_j)^{2n}
 =\half\sum_{i,j}(\xi_i-\xi_j)^{2n}
 =\half\sum_{i,j} \sum_{l=0}^{2n} (-1)^l{2n\choose l}\xi_i^l \xi_j^{2n-l}
 \notag\\
 &\to \half \sum_{l=0}^{2n} (-1)^l {2n\choose l}\tr X^l \tr X^{2n-l}
\notag\\
 &=\sum_{l=0}^{n-1} (-1)^l  {2n\choose l}\tr X^l \tr X^{2n-l}
 +{(-1)^n\over 2} {2n\choose n}(\tr X^n)^2\equiv (\Delta X)^{2n},
\end{align}
and use the contraction rule
\begin{align}
 \ev{X^\alpha{}_\beta X^\gamma{}_\delta}=f_2^{-1}\delta^\alpha_\delta \delta_\beta^\gamma .
\end{align}
Some of the correlators computed using the matrix model diagrams are:
\begin{align}
  \ev{\xi^2}&=\ev{\tr X^2}=N^2,\qquad
 \ev{(\xi^1)^2}=\ev{(\tr X)^2}=N,\nn
 \ev{(\Delta\xi)^2}&=\ev{N\tr X^2-(\tr X)^2}=N^3-N,\nn
 \ev{\xi^4}&=\ev{\tr X^4}=2N^3+N,\qquad
 \ev{\xi^3\xi^1}=\ev{\tr X^3 \tr X}=3N^2,\nn
 \ev{\xi^2\xi^2}&=\ev{(\tr X^2)^2}=N^4+2N^2,\qquad
 \ev{\xi^2 (\xi^1)^2}=\ev{\tr X^2 (\tr X)^2}=N^3+2N,\nn
 \ev{(\xi^1)^4}&=\ev{(\tr X)^4}=3N^2,\nn
 \ev{(\Delta \xi)^4}&=\ev{N\tr X^4-4\tr X^3 \tr X+3(\tr X^2)^2}=5N^4-5N^2,\nn
 \ev{(\Delta \xi^2)^2}&=\ev{[N\tr X^2-(\tr X)^2]^2}=N^6-N^2,\nn[1ex]
 \ev{\xi^6}&=\ev{\tr X^6}=5N^4+10N^2,\qquad
 \ev{\xi^4 \xi^2}=2N^5+9N^3+4N,\nn
 \ev{\xi^4 (\xi^1)^2}&=2N^4+13N^2,\qquad
 \ev{\xi^4 (\Delta \xi)^2}=\ev{\xi^4[N\tr X^2-(\tr X)^2]}=2N^6+7N^4-9N^2,\nn
 \ev{(\xi^4)^2}&=\ev{(\tr X^4)^2}=4 N^6 + 40 N^4 + 61 N^2.\label{mmcorrs}
\end{align}
In the above expressions, we set $f_2=1$ for simplicity, but the correct
powers of $f_2$ can be recovered on dimensional grounds.  When computing
correlators such as \eqref{mmcorrs}, diagrams get out of hand quickly as
the power grows.  Rather than directly dealing with diagrams, it is
easier to assume that a given correlator is an even/odd polynomial in
$N$ with certain degree, and determine the coefficients by computer for
some small values of $N$.

So, in terms of the Hermitian matrix model, the ``free energy''
$\cF(N;M)_k$ can be computed as follows:
\begin{align}
 \cF(N;M)_k
 &=-k N f_0+{N^2\over 2}\log {kf_2\over\pi }
 - {N\over 2}\log {2k^2\over \pi}-\log G_2(N+1)
 \nn
 &\qquad +
 \Ev{
 \exp\left[{\sum_{n=2}^\infty {(-1)^n f_{2n}\over (2n)!\, k^{2n-1}}\tr X^{2n}
 +
 \sum_{n=1}^\infty c_{2n} \Bigl({\pi^2 \over k}\Bigr)^{n} (\Delta X)^{2n}}\right]
 -1}_{\!\!\rm conn},
 \label{F_X_int_conn}
\end{align}
where $\ev{~}_{\rm conn}$ means the connected part; for example,
\begin{align}
 \ev{(\tr X^2)^2}_{\rm conn}
 =
 \ev{(\tr X^2)^2}-\ev{\tr X^2}^2.
\end{align}
Carrying out the diagram expansion in \eqref{F_X_int_conn} to a few
orders and using the large $k$ expansion of $f_{2n}(k,t)$ given in
\eqref{f_n}, we obtain the following large $k$ expansion for $\cF(N;M)_k$:
\begin{align}
 &\cF(N;M)_k\notag\\
 &=
 {2k N\over \pi}\,\cI\Bigl({\pi t\over 2}\Bigr)
 +{N^2\over 2}\ln {4k\over \pi \sin(\pi t)}
 -{N\over 2}\ln{2k^2\over \pi} 
 -\ln G_2(N+1)\nn
&\quad
-\frac{\pi N \left(2 N^2-1\right) }{48 \sin (\pi  t )\, k} \left[3 \cos (2 \pi  t )+1\right]
 \nn
&\quad
-\frac{\pi ^2 N^2  }
 {2304 \sin ^2(\pi  t) \, k^2}
 \left[
 (17 N^2+1) \cos (4 \pi  t )
 +4(11 N^2-29) \cos (2 \pi  t ) 
 -157   N^2+211
 \right]
 \nn
&\quad
-\frac{\pi ^3  N}{552960  \sin^3 (\pi  t )  k^3}
\Bigl[
 (674 N^4+250 N^2+201) \cos (6 \pi  t )
 -6 (442 N^4+690 N^2-427) \cos (4 \pi  t ) \notag\\ &\qquad\qquad
   +3 (2282N^4+3490 N^2-3635)\cos (2 \pi  t )
 +4348N^4-21940 N^2+12750
 \Bigr]
\notag\\
&\quad
-
\frac{\pi ^4  N^2 }{22118400 \sin ^4(\pi  t )k^4}
   \Bigl[
 (6223  N^4+8330 N^2+2997) \cos (8 \pi  t )\notag\\ &\qquad\qquad
 -8 (3983 N^4+6730 N^2-363) \cos (6 \pi  t )
 +20 (3797 N^4+1870 N^2+1623) \cos (4 \pi  t )\notag\\ &\qquad\qquad
 -8 (22249 N^4-44410 N^2+37011) \cos (2 \pi  t)
 -56627 N^4+113630 N^2-18753
 \Bigr]
\notag\\
 &\quad +\cO(k^{-5}).
 \label{F_expl_k}
\end{align}
Rewriting this as a large $M$ expansion gives Eq.~\eqref{F_expl}
presented in the main text.

%%%%%%%%%%%%%%%%%%%%%%%%%%%%%%%%%%%%%%%%%%%
%%%        MAIN TEXT ENDS HERE
%%%%%%%%%%%%%%%%%%%%%%%%%%%%%%%%%%%%%%%%%%%

\end{document}